\documentclass[12pt,preprint]{aastex}

\newcommand{\tpow}[1]{\cdot 10^{#1}}

\begin{document}


\title{Parameterization of $\gamma$, $e^{\pm}$ and Neutrino 
Spectra Produced by $p-p$ Interaction in Astronomical Environment}

\author{Tuneyoshi Kamae\altaffilmark{1},
Niklas Karlsson\altaffilmark{2},
Tsunefumi Mizuno\altaffilmark{3},
Toshinori Abe\altaffilmark{4},
Tatsumi Koi}
\affil{Stanford Linear Accelerator Center, Menlo Park, CA 94025}
\email{kamae@slac.stanford.edu}

\altaffiltext{1}{Also with Kavli Institute for Particle Astrophysics and Cosmology, Stanford University,
Menlo Park, CA 94025}
\altaffiltext{2}{Visiting scientist from Royal Institute of Technology, SE-10044 Stockholm, Sweden} 
\altaffiltext{3}{Present address: Department of Physics, Hiroshima University, Higashi-Hiroshima, Japan
739-8511} 
\altaffiltext{4}{Present address: Department of Physics, University of Tokyo, Tokyo, Japan 113-0033}

\begin{abstract}
We present the yield and spectra of stable secondary particles 
($\gamma$, $e^{\pm}$, $\nu_{e}$, $\bar{\nu}_{e}$, 
$\nu_{\mu}$, and $\bar{\nu}_{\mu}$) of $p-p$ interaction in 
parameterized formulae to facilitate calculations involving 
them in astronomical environments.
The formulae are derived from the up-to-date 
$p-p$ interaction model by \citet{Kamae05}, which incorporates 
the logarithmically rising inelastic cross section, the diffraction 
dissociation process, and the Feynman scaling violation.
To improve fidelity to experimental data in lower energies,
two baryon resonance contributions have been added:
one representing $\Delta$(1232) and the other representing multiple resonances
around 1600 MeV/c$^2$.  The parametrized formulae predict that all 
secondary particle spectra be harder by about 0.05 in power-law 
indices than that of the incident proton and their inclusive 
cross-sections be larger than those predicted by $p-p$ interaction 
models based on the Feynman scaling. 
\end{abstract}



\section{Introduction}

Gamma-ray emission due to neutral pions produced by
proton-proton interaction has long been predicted from the Galactic ridge, 
supernova remnants (SNRs), active galactic nucleus (AGN) jets, 
and other astronomical sites \citep{Hayakawa69, Stecker71, 
MurthyWolfendale86, Schoenfelder01, Schlickeiser02, Aharonian04a, 
AharonianVolkHorns04}.  High energy neutrinos 
produced by $p-p$ interaction in AGN jets will soon be detected 
with large-scale neutrino detectors that are under construction \citep{Halzen05}.
Spectra of these gamma-rays, neutrinos, and other secondaries depend heavily 
on the incident proton spectrum, which is unknown and needs to be derived, 
in almost all cases, from the observed spectra themselves. Such analyses
often involve iterative calculations with many trial proton spectra. 
The parameterized model presented here is aimed to improve accuracy
of such calculations.

Among the secondaries of $p-p$ interaction in astronomical environment, 
gamma-rays have been best studied. The p-p interaction is one of the two
dominant gamma-ray emission mechanisms in the sub-GeV 
to multi-TeV range, the other being Compton up-scattering of 
low energy photons by high energy electrons.  
Gamma-rays in this energy range have been detected from pulsars, 
the Galactic Ridge, SNRs, blazars and other source categories
\citep{Stecker71, MurthyWolfendale86, Ong98, Schoenfelder01, Weekes03, Aharonian04a}. 

High energy gamma-rays from AGN jets are interpreted as mostly due to 
the inverse Compton up-scattering of low-energy 
photons by multi-TeV electrons. Observed radio and X-ray spectra 
match those of synchrotron radiation by these electrons. 
Synchronicity in variability between the observed X-ray and gamma-ray 
fluxes has given strong support for the inverse Compton 
up-scattering scenario \citep{Ong98, Schoenfelder01, Schlickeiser02, 
Aharonian04a}. For some AGN jets, the above scenario 
does not work well, and $p-p$ interaction has been proposed 
as an alternative mechanism 
\citep{MueckeProtheroe01, Muecke03, BoettcherReimer04}.

High energy gamma-rays detected by COS-B and EGRET from the Galactic 
Ridge, on the other hand, are interpreted 
as predominantly due to neutral pions produced by interaction of 
protons and nuclei with the interstellar matter
\citep[ISM;][]{Stecker73, Strong78, Strong82, SMR00, SB81, Hasselwander82, Bloemen84, 
Bloemen85, Dermer86a, Stecker90, Hunter97, Mori97, Stanev04}. 
The measured gamma-ray flux and spectral 
shape \citep{Hunter97} have been viewed as the key attestation 
to this interpretation.
It is also known that inverse Compton scattering contributes significantly
to the Galactic Ridge gamma-ray emission \citep{MurthyWolfendale86, 
SMR00, Schoenfelder01}.

In the past two years, several SNRs, including RX J1713-3946 
and RX J0852-4622, have been imaged in the TeV band with an angular 
resolution around $0.1^{\circ}$ by H.E.S.S. \citep{Aharonian04b, 
Aharonian04c, Aharonian05}. A smooth featureless spectrum, suggestive of 
synchrotron radiation by multi-TeV electrons, has been detected 
in the X-ray band from RX J1713-3946 \citep{Koyama97, Slane99, Uchiyama03}
and RX J0852-4622 \citep{Tsunemi00, Iyudin05}.
The measured TeV gamma-ray fluxes and spectra, however, do not agree 
well with those predicted by the inverse Compton scenario
(see, e.g., the analysis in \citet{Uchiyama03}). Several authors 
have proposed that the TeV gamma-rays are possibly due to interaction of 
accelerated protons with the ISM \citep{BerezhkoVolk00, Enomoto02, 
Aharonian04a, Katagiri05}. 

Higher precision data are expected, in the GeV range, 
from GLAST Large Area Telescope (GLAST-LAT 2005)\footnote{GLAST Large
Area Telescope, http://www-glast.stanford.edu.} and, in the TeV range, 
from the upgraded Air Cherenkov Telescopes \citep{AharonianVolkHorns04}:
they will soon test applicability of 
the inverse Compton up-scattering and the proton interaction with ISM
for various astronomical gamma-ray sources.
In many objects, secondary electrons and positrons may produce
fluxes of hard X-rays and low-energy gamma-rays detectable with 
high-sensitivity instruments aboard \emph{Integral}, \emph{Swift},
\emph{Suzaku}, and \emph{NuSTAR}.\footnote{See \emph{INTEGRAL} Web site,
http://sci.esa.int/esaMI/Integral, the \emph{NuSTAR} Web site,
http://www.nustar.caltech.edu, the \emph{Suzaku} Web site
http://www.isas.jaxa.jp/e/enterp/missions/astro-e2, and the \emph{Swift}
Web site, http://swift.gsfc.nasa.gov/docs/swift.} 
The formulae given here will give fluxes and 
spectra of these secondary particles for arbitrary incident proton 
spectrum.

This work is an extension of that by \citet{Kamae05}, 
where up-to-date knowledge of the $\pi^{0}$ yield in the
$p-p$ inelastic interaction has been used to predict the Galactic 
diffuse gamma-ray emission. The authors have found that past 
calculations \citep{Stecker70, Stecker73, Stecker90, Strong78, SB81,
Dermer86a, Dermer86b, Mori97, SMR00} 
had left out the diffractive interaction and 
the Feynman scaling violation in the non-diffractive inelastic 
interaction.  Another important finding by them is that most 
previous calculations have assumed 
an energy-independent $p-p$ inelastic cross-section of about 24 mb 
for $T_{p}\gg 10$ GeV, whereas recent experimental data have 
established a logarithmic increase with the incident proton energy. 
Updating these shortfalls has changed the
prediction on the gamma-ray spectrum in the GeV band significantly:
the gamma-ray power-law index is harder 
than that of the incident proton; and the GeV$-$TeV 
gamma-ray flux is significantly larger than that predicted 
on the constant cross-section and Feynman scaling \citep{Kamae05}. 
The model by \citet{Kamae05} will hereafter be referred to as model A.

Model A does not model $p-p$ interaction accurately 
near the pion production threshold.  To improve prediction of
gamma-rays, electrons, and positrons produced near the pion 
production threshold, two baryon resonance 
excitation components have bee added to model A: $\Delta$(1232),
representing the $\Delta$ resonance, and res(1600), representing
resonances around 1600 MeV/c$^2$. 
We note here that $\Delta$(1232) is the most prominent and lightest 
Baryon resonance excited in $p-p$ interaction. It has a
mass of 1.232 GeV/c$^2$ and a width of about 0.12 GeV/c$^2$, and 
decays to a nucleon (proton or neutron) and a pion ($\pi^{+,0,-}$).
The other resonance, res(1600), is assumed to decay to a nucleon
and two pions. Introduction of these contributions have
necessitated adjustment of  the model A at lower energies as
described below.  The readjusted model will be referred to as the
``readjusted model A''.

The parameterized model presented here exhibits all features of 
model A at higher energies (proton kinetic energy, $T_{p}> 3$ GeV) 
and reproduces experimental data down to the pion 
production threshold.
The inclusive gamma-ray and neutrino 
cross-section formulae can be used to predict their yields and spectra 
for a wide range of incident proton spectrum. Formulae for electrons and 
positrons predict sub-TeV to multi-TeV 
secondary electrons and positrons supplied by $p-p$ interaction.  
We note that space-borne 
experiments such as PAMELA\footnote{See http://wizard.roma2.infn.it/pamela.}
will soon measure 
the electron and positron spectra in the sub-TeV energy range, 
where the secondaries of $p-p$ interaction
may become comparable to the primary components
\citep{Mueller01, Stephens01, HEAT01}. 

Due to paucity of experimental data and widely accepted
modeling, we have not parameterized inclusive secondary cross-sections 
for $\alpha - p$ nor $p -$He nor $\alpha -$He interactions. 
We note that $\alpha$-particles are 
known to make up about 7\% by number of cosmic-rays observed
near the Earth \citep{Schlickeiser02} and
that He to make up $\sim 10$\% by number of interstellar gas.
The total non $p-p$ contribution is comparable to that of $p-p$
contribution.  The $\alpha$-particle and He nucleus can be 
regarded, to a good approximation, as four independent protons 
beyond the resonance region ($T_p > 3$~GeV):
the error introduced is expected to be less than 10\%
for high energy light secondary particles \citep{Kamae05}.
  
Inclusion of $\alpha$-particle as projectile and He nuclei 
as targets will change the positron-electron ratio 
significantly (about $10-15$\%) as discussed below.
Fermi motion of nucleons and multiple nucleonic interactions 
in the nucleus are known to significantly affect pion production
near the threshold and in the resonance region 
\citep[$T_p < 3$~GeV;][]{Crawford80,Martensson00};
we acknowledge need for separate treatment of $p -$He, 
$\alpha -p$, and $\alpha -$He interactions in the future. 

\section{Monte Carlo Event Generation}

The parameterization of the inclusive cross sections
for $\gamma$, $e^{\pm}$, $\nu$, and $\bar{\nu}$ 
has been carried out, separately, for non-diffractive, 
diffractive, and resonance-excitation processes, in three steps: 
First, the secondary particle spectra have been extracted out of
events generated for mono-energetic protons 
(0.488 GeV $< T_{p} <$ 512 TeV) based on the readjusted model A. 
We then fit these spectra with a common parameterized 
function, separately 
for non-diffractive, diffractive and resonance-excitation processes. 
In the third step, the parameters determined for mono-energetic 
protons are fitted as functions of proton energy, again 
separately for the three processes. 
The above procedure has been repeated for all secondary particle
types.

The functional formulae often introduce tails extending beyond
the energy-momentum conservation limits, which may produce artifacts 
when wide range spectral energy density  
($E^2 d{\rm{flux}}(\gamma)/dE$) is plotted. 
To eliminate such artifacts, we introduce another set of functions 
to impose the kinematic limits. 

Several simulation programs have been used 
in model A \citep{Kamae05}: for the high energy non-diffractive process 
($T_{p} > 52.6$ GeV), Pythia 6.2 \citep{Pythia62} 
with the multi-parton-level scaling violation option
\citep{Sjostrand04};\footnote{See http://cepa.fnal.gov/CPD/MCTuning1 and
http://www.phys.ufl.edu/$\sim$rfield/cdf.} for the lower energy non-diffractive process,
the parameterized model by \citep{Blattnig00}; 
for the diffractive part, the program by 
T. Kamae (2004, personal communication)\footnote{Diffractive process 
has been included in Pythia after the work began.}. 
In the readjusted model A, two programs to simulate two
resonance-excitation components have also been added. 
Modeling of the two resonance components will be explained below. 

\section{Non-Diffractive, Diffractive, and Resonance-Excitation Cross-Sections}

\begin{figure}
\begin{center}
\scalebox{0.75}{\plotone{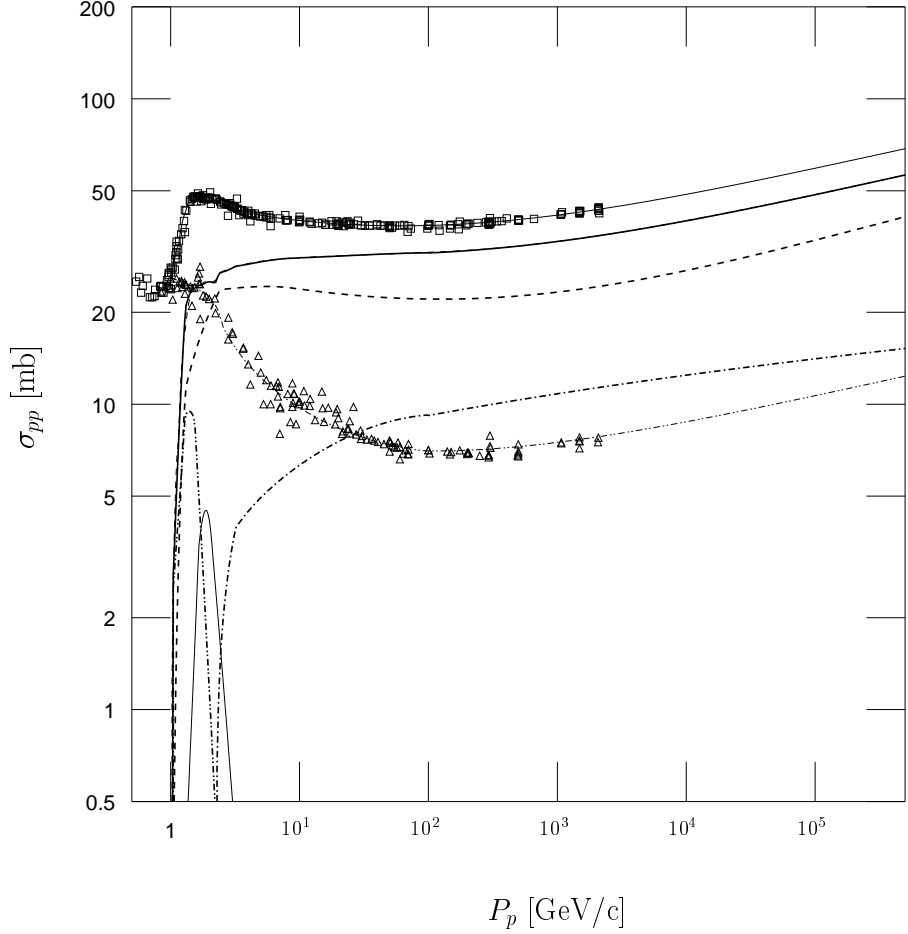}}
\end{center}
\caption{Experimental $p-p$ cross sections, as a function of proton momentum,
and that of readjusted model A: experimental total (squares),
experimental elastic (triangles),
total inelastic (thick solid line), non-diffractive (dashed line),
diffractive process (dot-dashed line), $\Delta$(1232) (dotted line), 
and res(1600) (thin solid line).
The total inelastic is the sum of the four components.
The thin solid and dot-dot-dashed lines running through the two 
experimental data sets are eye-ball fits to the total and 
elastic cross sections, respectively.}
\label{fig:cross-sectionsA}
\end{figure}

Experimental data on $p-p$ cross-sections are archived 
for a broad range of the incident proton energy and various final
states.  The total and elastic cross-sections have been compiled 
from those by \citet{Hagiwara02}, as shown in 
Figure \ref{fig:cross-sectionsA}.
The two thin curves running through experimental data points 
in the figure are our eye-ball fits to
the total and elastic cross sections.
We then define the ``empirical'' inelastic cross section
as the difference of the two curves to which the sum of 
non-diffractive, diffractive, $\Delta$(1232)-excitation,
and res(1600)-excitation components are constrained.
Typical errors in the empirical inelastic cross-section are
20\% for $T_{p} < 3$ GeV and 10\% for $T_{p} > 3$ GeV.

\begin{figure}
\begin{center}
\scalebox{0.75}{\plotone{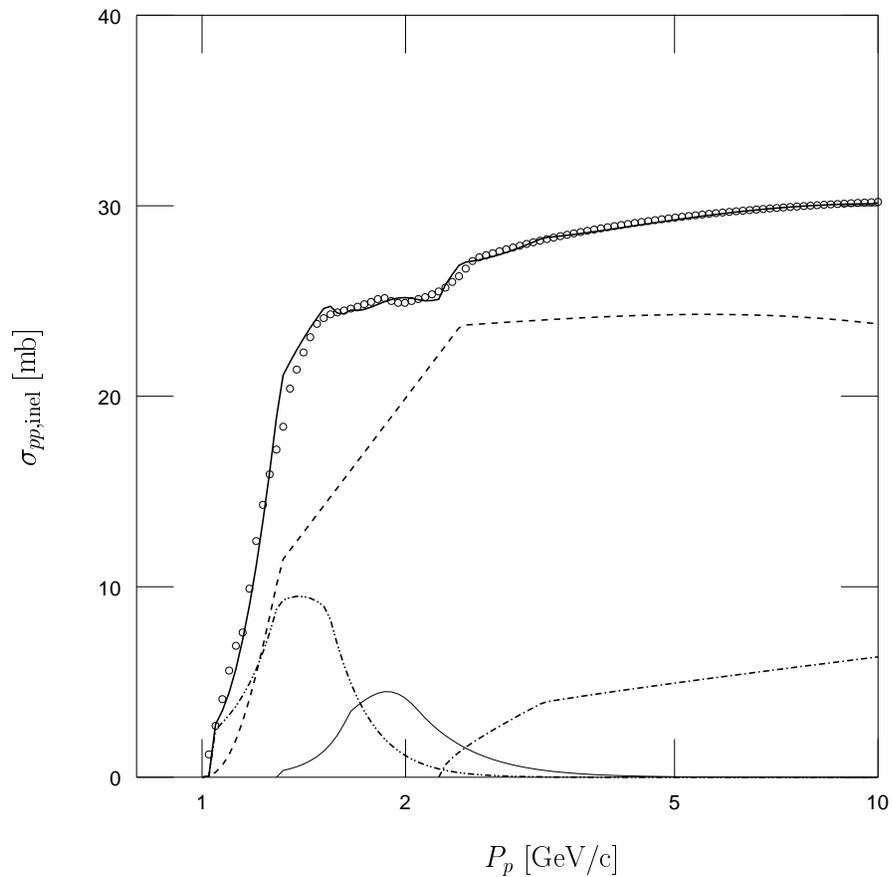}}
\end{center}
\caption{Experimental $p-p$ cross sections, as a function of proton momentum,
and that of readjusted model A for $T_p<10$ GeV. Small circles represent the
empirical inelastic cross section described in the text. Lines are the same
as in Figure \ref{fig:cross-sectionsA}.}
\label{fig:cross-sectionsB}
\end{figure}

The four component cross sections of the readjusted model A
and their sum are shown in 
Figures \ref{fig:cross-sectionsA} and \ref{fig:cross-sectionsB}. 
The empirical inelastic cross section is shown by a series 
of small circles in Figure \ref{fig:cross-sectionsB}.
The component cross sections take formulae given 
in equations (\ref{eq:sigmaNonDiff}) (non-diffractive), 
(\ref{eq:sigmaTotDiff}) (diffractive), 
(\ref{eq:sigmaDelta}) [$\Delta$(1232)] and (\ref{eq:sigmaReson})
[res(1600)]. These are also shown 
in Figures \ref{fig:cross-sectionsA} and \ref{fig:cross-sectionsB}.

We note that there is no clear experimental method separating 
the four components, especially at lower energies ($T_{p} < 20$ GeV). 
This ambiguity does not significantly affect the secondary particle 
fluxes, as long as the sum agrees with the total 
inelastic cross section and the total secondary inclusive cross sections
agree with the corresponding experimental data.

The secondary particle spectra for a mono-energetic proton are normalized
to the component cross sections given in equations (\ref{eq:sigmaNonDiff}), 
(\ref{eq:sigmaTotDiff}), (\ref{eq:sigmaDelta})
and (\ref{eq:sigmaReson}) at the corresponding proton energy.
We note that the non-diffractive component for $T_{p} < 52.6$ GeV
is based on the formula by \citet{Blattnig00},
which is normalized to their $\pi^{0}$ inclusive cross section
formula not to the total inelastic cross section.
In the readjusted model A, this component cross section 
is defined by equation (\ref{eq:sigmaNonDiff})
and the $\pi^{0}$ inclusive cross section formula of \citet{Blattnig00}
has been redefined so that the sum of the four components
reproduce the experimental $\pi^{0}$ inclusive cross section.
The positive and negative pion inclusive cross sections
are also redefined as products of our $\pi^{0}$ inclusive cross section
and the ratio of the $\pi^{+,-}$ and $\pi^{0}$ inclusive cross sections 
given in \citet{Blattnig00}. 
\clearpage

\clearpage
\begin{equation}\label{eq:sigmaNonDiff}
\sigma_{\rm{NonDiff}}^{pp}(x)\rm{[mb]} = \left\{ \begin{array}{ll}
0 & P_{p} < 1\ \rm{GeV/c},\\
0.57(x/a_{0})^{1.2}(a_{2} + a_{3}x^{2} + a_{4}x^{3}\\
\ \ + a_{5}\exp(-a_{6}(x + a_{7})^{2})) & 1\leq P_{p} \leq 1.3\ \rm{GeV/c},\\
(b_{0}|a_{1} - x| + b_{1}|a_{0} - x|)/(a_{1} - a_{0}) & 1.3\leq P_{p} \leq 2.4 \rm{GeV/c},\\
a_{2} + a_{3}x^{2} + a_{4}x^{3}\\
\ \ + a_{5}\exp(-a_{6}(x + a_{7})^{2}) & 2.4\leq P_{p} \leq 10\ \rm{GeV/c},\\
c_{0} + c_{1}x + c_{2}x^{2} & P_{p} > 10\ \rm{GeV/c}, 
\end{array}\right.
\end{equation}

\begin{equation}\label{eq:sigmaTotDiff}
\sigma_{\rm{Diff}}^{pp}(x)[\rm{mb}] = \left\{ \begin{array}{ll}
0 & P_{p}<2.25\ \rm{GeV/c},\\
\sqrt{(x - d_{0})/d_{1}}\\
\ \ \times(d_{2} + d_{3}\log_{10}(d_{4}(x - 0.25))\\
\ \ + d_{5}x^{2} - d_{6}x^{3}) & 2.25 \leq P_{p} \leq 3.2\ \rm{GeV/c},\\
d_{2} + d_{3}\log_{10}(d_{4}(x - 0.25))\\
\ \ + d_{5}x^{2} - d_{6}x^{3} & 3.2 \leq P_{p} \leq 100\ \rm{GeV/c},\\
e_{0} + e_{1}x & P_{p} > 100\ \rm{GeV/c},
\end{array}\right.
\end{equation}

\begin{equation}\label{eq:sigmaDelta}
\sigma_{\rm{\Delta 1232}}^{pp}(x)[\rm{mb}] = \left\{ \begin{array}{ll}
0 & E_{p}<1.4\ \rm{GeV},\\
f_{0}E_{p}^{10} & 1.4 \leq E_{p} \leq 1.6\ \rm{GeV},\\
f_{1}\exp{(-f_{2}(E_{p} - f_{3})^{2})} & 1.6 \leq E_{p} \leq 1.8\ \rm{GeV},\\
f_{4}E_{p}^{-10} & 1.8 \leq E_{p} \leq 10\ \rm{GeV},\\
0 & E_{p} > 10\ \rm{GeV},
\end{array}\right.
\end{equation}

\begin{equation}\label{eq:sigmaReson}
\sigma_{\rm{Res(1600)}}^{pp}(x)[\rm{mb}] = \left\{ \begin{array}{ll}
0 & E_{p}<1.6\ \rm{GeV},\\
g_{0}E_{p}^{14} & 1.6 \leq E_{p} \leq 1.9\ \rm{GeV},\\
g_{1}\exp{(-g_{2}(E_{p} - g_{3})^{2})} & 1.9 \leq E_{p} \leq 2.3\ \rm{GeV},\\
g_{4}E_{p}^{-6} & 2.3 \leq E_{p} \leq 20\ \rm{GeV},\\
0 & E_{p} > 20\ \rm{GeV},
\end{array}\right.
\end{equation}

where $x=\log_{10}(P_{p}{\rm{[GeV/c]}})$ and $E_{p}$ is the proton energy in GeV.

\subsection{Introduction of Resonance-Excitation Processes to Model A}

One or both of the projectile and target protons can be
excited to baryon resonances in the $p-p$ interaction.
Here we use ``baryon resonances'' to represent 
both nucleon resonances (iso-spin=$1/2$) and $\Delta$ 
resonances (iso-spin=$3/2$). These excitations
enhance the pion production (and hence secondary particle production)
near the inelastic threshold. The most prominent resonance
among them is $\Delta$(1232), which has a mass of 1232 MeV/c$^2$ 
and decays predominantly ($>99$\%) to a nucleon and a pion
\citep{Hagiwara02}.

\citet{Stecker70} proposed a cosmic gamma-ray model in which neutral pions 
are produced only through the $\Delta$(1232) excitation
for $T_p \leq 2.2$ GeV. The resonance is assumed to move only 
in the direction of the incident proton. At higher energies, another
process, the fireball process, sets in and produces pions with limited
transverse momenta.

\citet{Dermer86a} compared predictions of models on $\pi^{0}$ kinetic 
energy distribution in the proton-proton 
center-of-mass (CM) system with experiments and noted that
the model by \citet{Stecker70} reproduces data better 
than the scaling model by \citet{SB81} for $T_p < 3$ GeV. 
He proposed a cosmic gamma-ray production model that covers a wider 
energy range by connecting the two models in the energy range
$T_p = 3 - 7$ GeV.

Model A by \citet{Kamae05} has been constructed primarily for the
$p-p$ inelastic interaction $T_p \gg 1$ GeV and has left
room for improvement for $T_p < 3$ GeV. The diffraction
dissociation component of model A has a resonance-excitation feature 
similar to that implemented in \citet{Stecker71} for $T_p > 3$ GeV
where either or both protons can be excited to nucleon resonances 
(iso-spin=1/2 and mass around 1600 MeV/c$^2$) along the 
direction of the incident and/or target protons.
What has not been implemented in model A is the enhancement
by baryon resonances
in the inclusive pion production cross sections below $T_p < 3$ GeV.

We note here that the the models by \citet{Stecker70} and 
by \citet[see also \citealt{Dermer86b}]{Dermer86a} used experimental 
data on the inclusive $\pi^{0}$ yield (and that of charged pions) to guide
their modeling, but not the total inelastic cross section. 
Model A by \citet{Kamae05}, on the other hand, has simulated 
all particles in each event (referred to as the
``exclusive'' particle distribution) for all component 
cross sections.  One exception is simulation of the low-energy 
non-diffractive process ($ T_p < 52.6$ GeV) by \citet{Blattnig00}. 
The inclusive $\pi^{0}$ (or gamma-ray) yield is 
obtained by collecting $\pi^{0}$ (or gamma-rays) in simulated 
exclusive events. 
When readjusting model A by adding the resonance-excitation feature 
similar to that by \citet{Stecker70}, overall coherence to model A 
has been kept.
We adjusted the $\Delta$(1232) excitation cross section to
reproduce the total inelastic cross section given in 
Figure \ref{fig:cross-sectionsB} and fixed
the average pion multiplicities
for $+:0:-$ to those expected by the one-pion-exchange hypothesis, 
$0.73:0.27:0.0$.  As higher-mass resonances begin to 
contribute, the average pion multiplicity is expected to
increase.  To reproduce the experimental $\pi^{0}$ inclusive 
cross section and total inelastic cross section for $T_p < 3$ GeV, 
we introduced a second resonance, res(1600).  This resonance does 
not correspond to any specific resonance but represents 
several baryon resonances at around $1600$ MeV/c$^2$: 
its pion multiplicities ($+:0:-$) are assumed to be 1.0:0.8:0.2.
We note here that the resonance components favor
positive pions significantly over neutral pions while negative
pions are strongly suppressed.

The distribution of pion kinetic energy in the $p-p$ center-of-mass 
(CM) system ($T_{\pi}$) has been adjusted to reproduce the experimental 
ones given in Figs. 3-5 of \citet{Dermer86a}.  
For the $\Delta$(1232) excitation, the probability increases  
proportionally to $T_{\pi}$ up to its maximum, set at 
$T_{\pi}=0.28\times {\rm{abs}}(T_{p}-0.4)^{0.45}$.  Here $T_{\pi}$ and 
$T_p$ are measured in GeV/c.  The distribution goes 
to zero beyond this maximum value.
For res(1600), the probability distribution increases proportionally
to $T_{\pi}$ to reach its peak at 
$T_\pi=0.16\times {\rm{abs}}(T_{p}-0.4)^{0.45}$.  It decreases linearly
until reaching zero at twice the peak of $T_{\pi}$.  

Pion momentum is directed isotropically in the $p-p$ CM system for 
res(1600) as well as for $\Delta$(1232). No angular correlation 
has been assumed between the two pions from res(1600): this is 
justified for the astronomical environment where chance of detecting
two gamma rays from a same interaction is null. Decay kinematics
including the polarization effect has been implemented
to the charged pion decay. This treatment allows
the resonances to recoil transversely to the direction of 
the incident proton while the recoil was constrained 
along the incident proton direction in \citet{Stecker70}.

\begin{figure}
\begin{center}
\scalebox{0.75}{\plotone{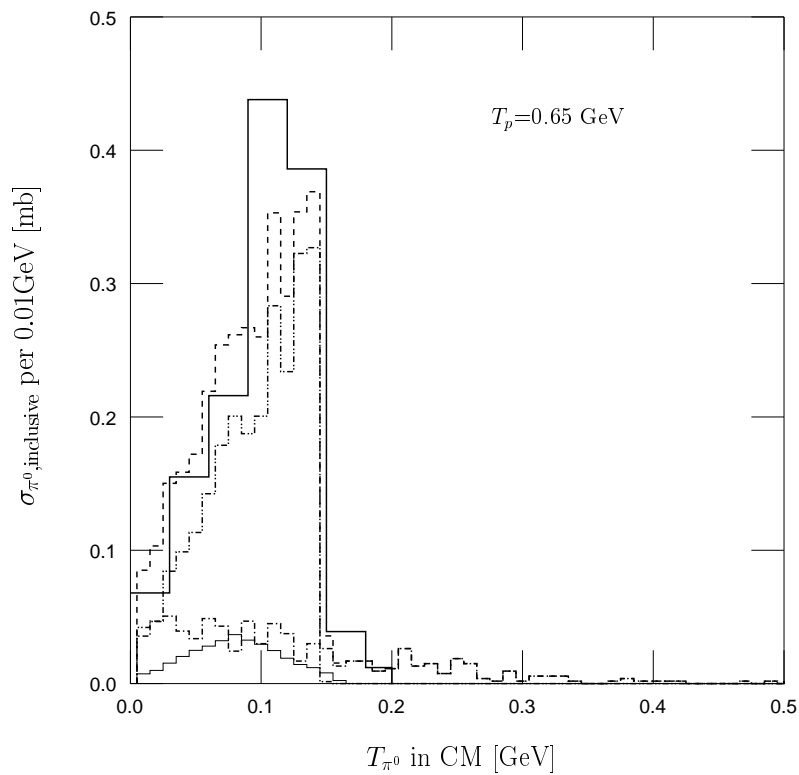}}
\end{center}
\caption{Experimental and simulated $\pi^{0}$ kinetic energy 
distributions in the $p-p$ CM system for $T_{p}=0.65$ GeV. 
The thick solid line is the experimental data taken from Fig. 3
of \citet{Dermer86a}. The dashed line is the sum of all readjusted
model A components: non-diffractive (dot-dashed line),
$\Delta$(1232) (double-dot-dashed line), and res(1600) (thin solid line).}
\label{fig:KEpion-distributionA}
\end{figure}

\begin{figure}
\begin{center}
\scalebox{0.75}{\plotone{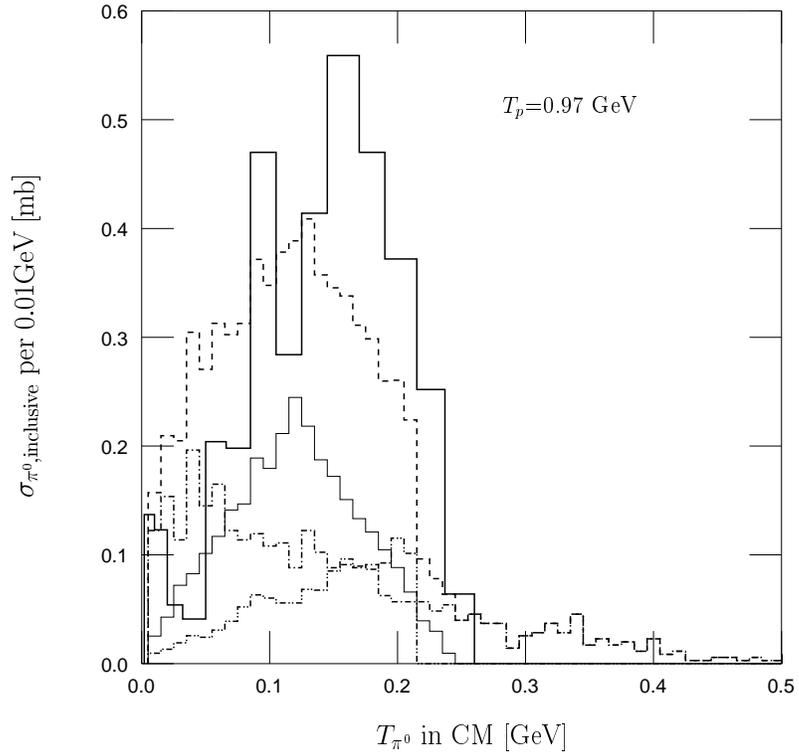}}
\end{center}
\caption{Experimental and simulated $\pi^{0}$ kinetic energy 
distributions in the $p-p$ CM system for $T_{p}=0.97$ GeV. 
The thick solid line represents experimental data (taken from Fig. 4
of \citet{Dermer86a}). Other lines are the same as 
in Fig. \ref{fig:KEpion-distributionA}.}
\label{fig:KEpion-distributionB}
\end{figure}

\begin{figure}
\begin{center}
\scalebox{0.75}{\plotone{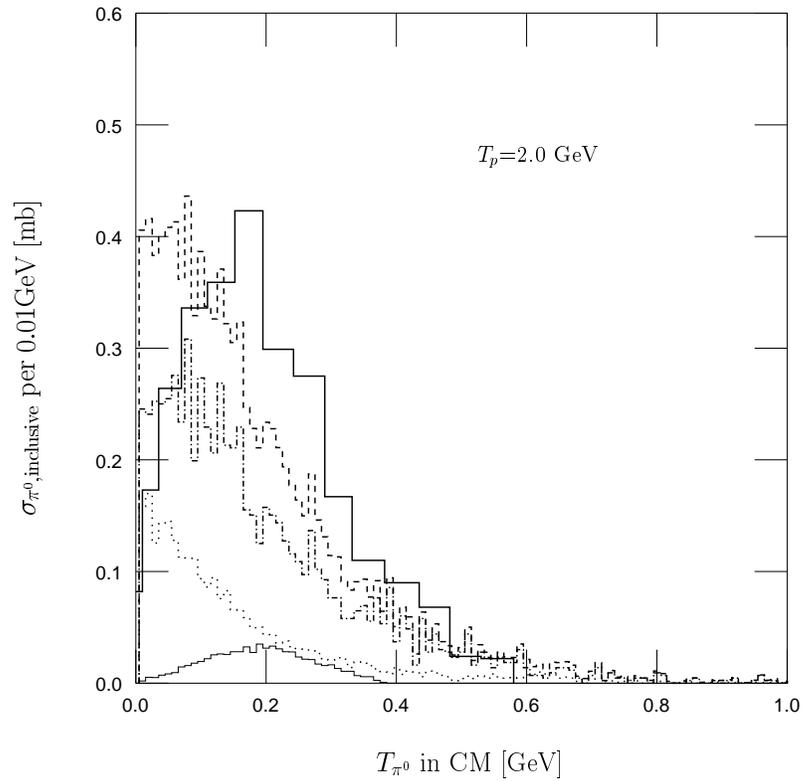}}
\end{center}
\caption{Experimental and simulated $\pi^{0}$ kinetic energy 
distribution in the $p-p$ CM system for $T_{p}=2.0$ GeV.
the thick solid line represents experimental data (taken from Fig. 5
of \citet{Dermer86a}), and the dotted line 
is for the diffractive process. The others are
the same as in Fig. \ref{fig:KEpion-distributionA}.}
\label{fig:KEpion-distributionC}
\end{figure}

To validate the resonance components of the readjusted model A,
we have compared the model $\pi^{0}$ spectrum in the 
$p-p$ CM system at $T_{p}=$ 0.65, 0.97, and 2.0 GeV with experimental
data in Figures \ref{fig:KEpion-distributionA}, \ref{fig:KEpion-distributionB},
\ref{fig:KEpion-distributionC}. Shown in these figures are
contributions of the $\Delta$(1232), res(1600), non-diffractive,
and diffractive processes.
Our model reproduces well the shape of pion kinetic energy 
distribution at $T_{p}=$ 0.65 GeV but begins to concentrate
more towards zero kinetic energy than experimental data 
at $T_{p}=$ 0.97 and 2.0 GeV.  Fidelity to the experimental
data is much improved when compared with the model by 
\citet{SB81} but somewhat worse than the one by 
\citet{Stecker70} given in Figures 2-6 of \citet{Dermer86a}.
The difference among the models becomes less noticeable 
for pion decay products, gamma-rays, electrons, positrons, and neutrinos.

\begin{figure}
\begin{center}
\scalebox{0.75}{\plotone{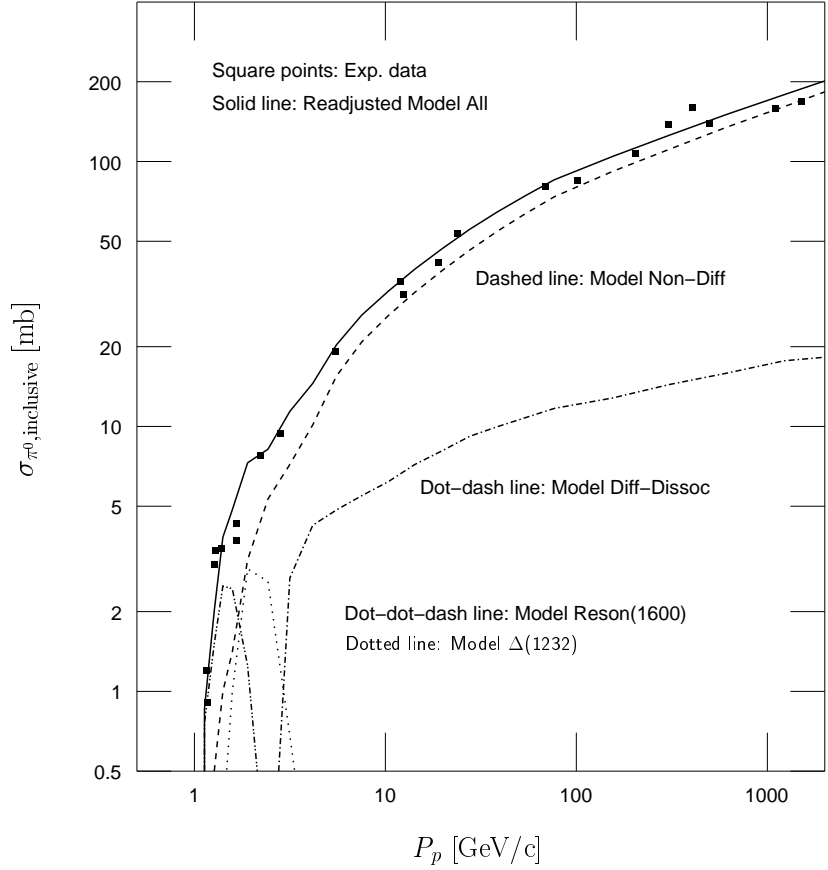}}
\end{center}
\caption{Experimental and simulated $\pi^{0}$ inclusive cross section.
Experimental data are those assembled by \citet{Stecker70} and
\citet{Dermer86a}. 
Lines are those of readjusted model A: total (solid line),
non-diffractive (dashed line),
diffractive process (dot-dashed line), $\Delta$(1232) (double-dot-dashed line),
and res(1600) (dotted line).}
\label{fig:NPi0ExpModel}
\end{figure}

Our inclusive $\pi^0$ cross section, sum of all four components,
is compared with experimental data 
assembled by \citet{Stecker70} and \citet{Dermer86a} 
in Fig. \ref{fig:NPi0ExpModel}. The readjusted model A
reproduces experimental data quite well for a wide range of
incident proton energy.  
\clearpage

\clearpage
\section{Inclusive Spectra of Simulated Events for Mono-Energetic Proton Beam}

The first step of parameterization is to generate simulated events
for mono-energetic protons. To simplify this step, events 
have been generated for discrete proton energies at a geometrical series of 
$T_{p}=1000.0 \times 2^{(N-22)/2}$ GeV where $N=0-40$. 
Each proton kinetic energy ($T_p$) represents a bin covering between 
$2^{-0.25}T_{p}$ and $2^{0.25}T_{p}$. The sampling density has been 
increased for $T_{p}<1$ GeV by adding points at $T_{p} =$ 0.58 GeV 
and 0.82 GeV.

\begin{figure}
\begin{center}
\scalebox{0.75}{\plotone{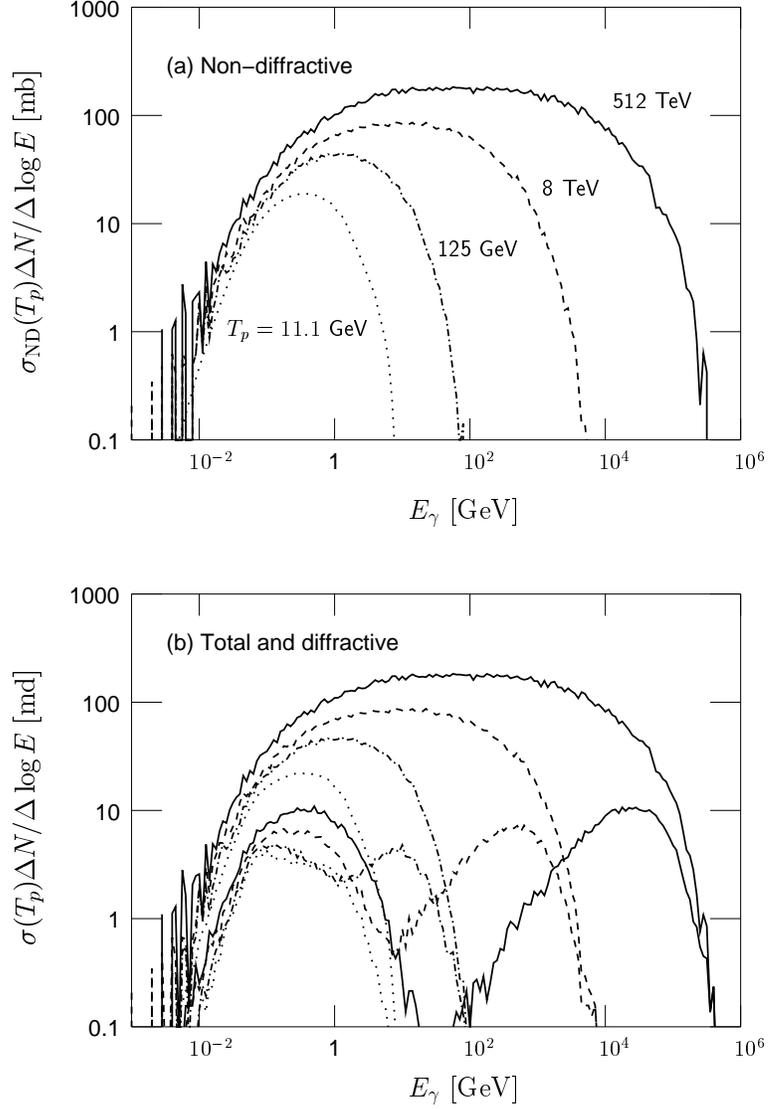}}
\end{center}
\caption{Gamma-ray inclusive cross sections by readjusted model A for 
four mono-energetic protons for
(a) the non-diffractive process and (b) the total (top four curves) and 
the diffractive process (bottom four curves). Note that the diffractive 
process produces two humps at $T_{p}=$ 512 TeV, 8 TeV and 125 GeV. 
The four proton kinetic energies are $T_{p}=$ 512 TeV (solid line), 8 TeV 
(dashed line), 125 GeV (dot-dashed line) 
and 11.1 GeV (dotted line).
The bin width is $\Delta {\rm{log}}(E) = 0.05$.
The large fluctuation in the curves at the highest and lowest ends
are due to statistics in the simulation.}
\label{fig:gamma-ray-spectra-A}
\end{figure}

\begin{figure}
\begin{center}
\scalebox{0.75}{\plotone{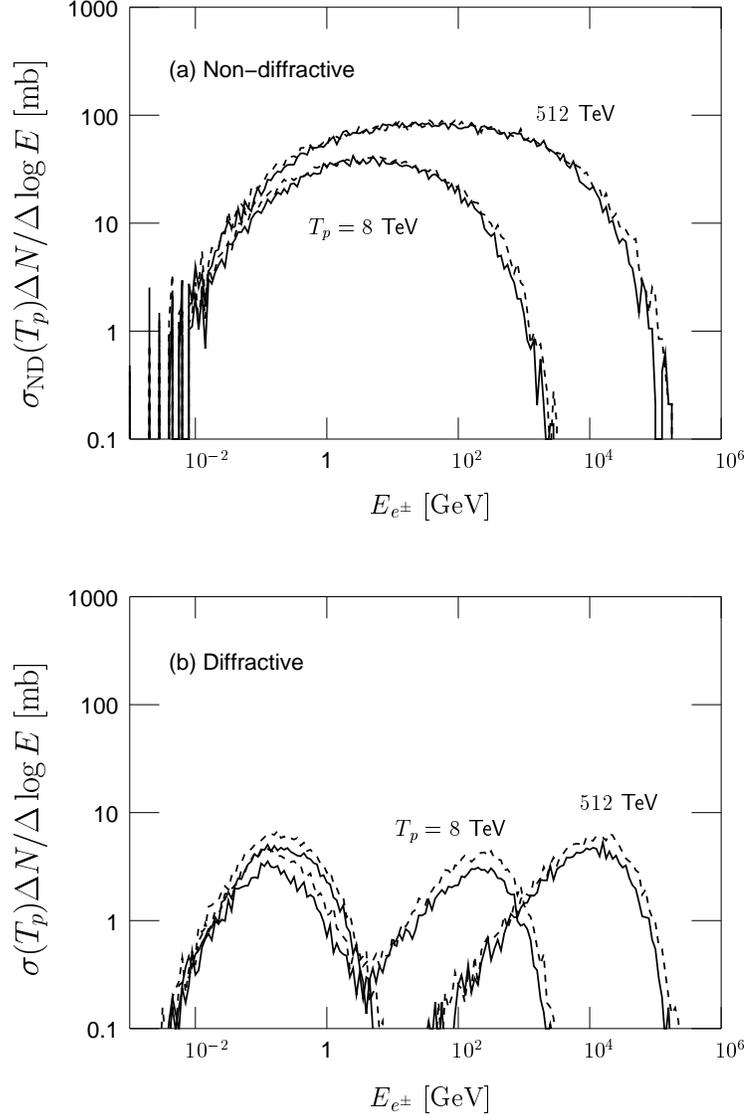}}
\end{center}
\caption{Electron and positron inclusive cross sections by 
readjusted model A 
for $T_{p}=$ 8 and 512 TeV for (a) the non-diffractive and (b)
diffractive processes. Electron spectra are shown by 
solid lines and positron spectra by dashed lines.
Note that the spectra have two humps in the diffractive process.
The bin width is $\Delta {\rm{log}}(E) = 0.05$.
The large fluctuation in the histograms at the highest and lowest ends
are due to statistics in the simulation.}
\label{fig:elec-posi-spectra-A} 
\end{figure}

Secondary particle spectra are histogrammed from these 
simulated events in energy bins of width $\Delta E/E = 5$~\%.
Figures \ref{fig:gamma-ray-spectra-A}a and \ref{fig:gamma-ray-spectra-A}b 
show thus-obtained inclusive gamma-ray 
cross sections for the non-diffractive and diffractive processes, 
for $T_{p}=$ 512 TeV, 8 TeV, 125 GeV 
and 11.1 GeV respectively. Those for $e^{\pm}$ are given for $T_{p}=$ 8 TeV 
and 512 TeV in Figures \ref{fig:elec-posi-spectra-A}a and 
\ref{fig:elec-posi-spectra-A}b.
\clearpage

\clearpage
We then define, after several iterations of fitting,
functional formulae that reproduce the secondary particle spectra 
for mono-energetic protons, for the non-diffractive, 
diffractive, and resonance-excitation processes separately. 
For the non-diffractive process, the differential
inclusive cross section ($\Delta \sigma_{\rm{ND}}$) to produce 
a secondary particle 
in a bin of width $\Delta E_{\rm{sec}}/E_{\rm{sec}} = 100$~\% 
centered at $E_{\rm{sec}}$ is given as

\begin{equation}\label{eq:nonDiff}
\frac{\Delta\sigma_{\rm{ND}}(E_{\rm{sec}})}{\Delta \log(E_{\rm{sec}})} = F_{\rm{ND}}(x)F_{\rm{ND,kl}}(x),
\end{equation}

where $E_{\rm{sec}}\rm{[GeV]}$ is the energy of the secondary particle 
and $x=\log_{10}(E_{\rm{sec}}{\rm{[GeV]}})$.
$F_{\rm{ND}}(x)$ is the formula representing the non-diffractive 
cross section, given in equation (\ref{eq:Fnd}) below, and $F_{\rm{ND,kl}}(x)$
is the formula to approximately enforce the energy-momentum conservation limits:

\begin{eqnarray}\label{eq:Fnd}
F_{\rm{ND}}(x) &=& 
a_{0}\exp(-a_{1}(x - a_{3} + a_{2}(x - a_{3})^{2})^{2}) + \nonumber\\
&&a_{4}\exp(-a_{5}(x - a_{8} + a_{6}(x - a_{8})^{2} + a_{7}(x-a_{8})^{3})^{2}),
\end{eqnarray}

\begin{equation}\label{eq:Fndkl}
F_{\rm{ND,kl}}(x) = \frac{1}{(\exp{(W_{\rm{ND,l}}(L_{\rm{min}} - x))} + 1)}\frac{1}{(\exp{(W_{\rm{ND,h}}(x - L_{\rm{max}}))} + 1)},
\end{equation}

where $L_{\rm{min}}$ and $L_{\rm{max}}$ are the lower and upper kinematic
limits imposed and $W_{\rm{ND,l}}$ and  $W_{\rm{ND,h}}$ are the widths of
the kinematic cut-offs; $L_{\rm{min}}=-2.6$ for all secondary particles,
and the other parameters are listed in Table \ref{table:E_max-Delta}.

For the diffractive process we use a similar function

\begin{equation}\label{eq:diffDissoc}
\frac{\Delta\sigma_{\rm{diff}}(E_{\rm{sec}})}{\Delta \log(E_{\rm{sec}})} = F_{\rm{diff}}(x)F_{\rm{kl}}(x) 
\end{equation}

where $E_{\rm{sec}}\rm{[GeV]}$ is the energy of the secondary particle and
$x=\log_{10}(E_{\rm{sec}}{\rm{[GeV]}})$; $F_{\rm{diff}}(x)$ represents
the diffractive cross section, given in equation (\ref{eq:Fdiff}) below, and
$F_{\rm{kl}}(x)$ enforces the energy-momentum conservation:

\begin{eqnarray}\label{eq:Fdiff}
F_{\rm{Diff}}(x) &=& b_{0}\exp(-b_{1}((x - b_{2})/(1 + b_{3}(x - b_{2})))^{2}) + \nonumber\\
&&b_{4}\exp(-b_{5}((x - b_{6})/(1 + b_{7}(x - b_{6})))^{2}),
\end{eqnarray}

\begin{equation}\label{eq:Fdiffkl}
F_{\rm{kl}}(x) = \frac{1}{\exp{(W_{\rm{diff}}(x - L_{\rm{max}}))} + 1},
\end{equation}

with $W_{\rm{diff}}=75$ and $L_{\rm{max}}=\log_{10}(T_{p}[GeV])$.

For the resonance-excitation processes [$\Delta$(1232) and res(1600)] we use the function,

\begin{equation}\label{eq:delta}
\frac{\Delta\sigma_{\rm{res}}(E_{\rm{sec}})}{\Delta \log(E_{\rm{sec}})}=F_{\rm{res}}(x)F_{\rm{kl}}(x), 
\end{equation}

where $E_{\rm{sec}}\rm{[GeV]}$ is the energy of the secondary particle and
$x=\log_{10}(E_{\rm{sec}}{\rm{[GeV]}})$, $F_{\rm{res}}(x)$ represents
the cross section, given in equation (\ref{eq:Fdelta}) below, and 
$F_{\rm{kl}}(x)$, which is the same as for
the diffraction process, enforces the energy-momentum conservation:

\begin{equation}\label{eq:Fdelta}
F_{\rm{res}}(x) = c_{0}\exp{(-c_{1}((x - c_{2})/(1 + c_{3}(x - c_{2}) + c_{4}(x - c_{2})^{2}))^{2})}.
\end{equation}

To ensure that the parameterized model reproduces 
the experimental $\pi^{0}$ multiplicity after the readjustment
in the resonance-excitation region of $T_p$,
we have renormalized the non-diffractive contribution
by multiplying it with a renormalization factor, $r(T_{p})$, given below,
to the final spectrum. Note that this readjustment does not affected
the diffractive process. 

\begin{equation}
r(T_{p})\simeq 1.01 {\rm{\; \; \; for \; \; \;}} T_{p}> 1.95 {\rm{ GeV}},
\end{equation}

\begin{equation}
r(y=\log_{10}(T_{p})) = 3.05\exp{(-107((y + 3.25)/(1 + 8.08(y + 3.25)))^{2})}
{\rm{\; \; for \; \;}} T_{p}\leq 1.95 {\rm{ GeV}}.
\end{equation}

For all other secondary particles, $r(y=\log_{10}(T_{p}))$ is found in Tables
\ref{table:elecParameters}-\ref{table:antinumuParameters}.

\begin{figure}
\begin{center}
\scalebox{0.65}{\plotone{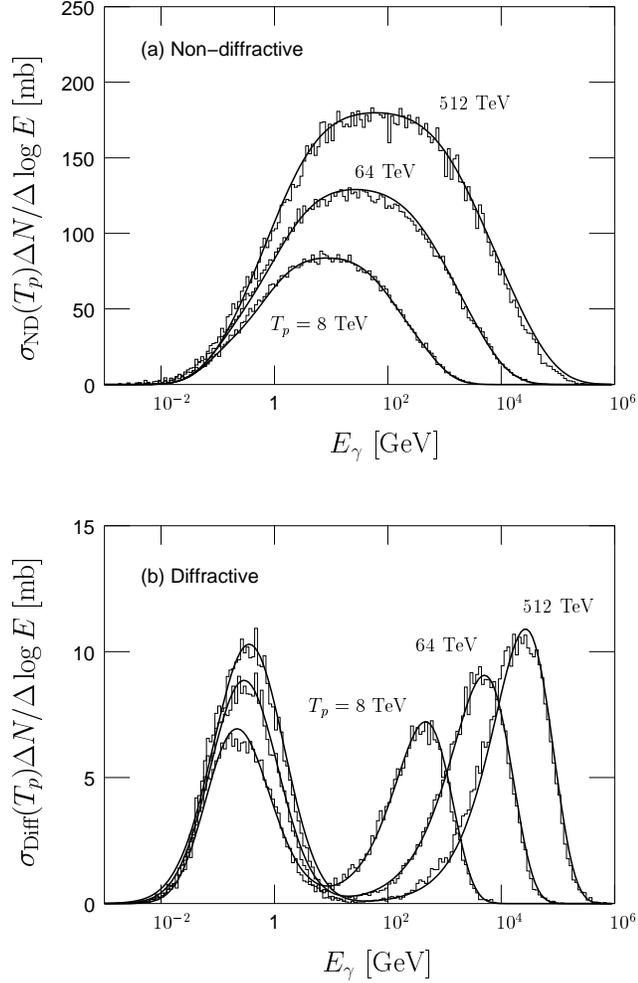}}
\end{center}
\caption{Simulated (histograms) and parameterized 
(solid lines) gamma-ray inclusive cross sections for 
(a) non-diffractive and (b) diffractive processes for 
$T_{p}=$ 8, 64 and 512 TeV. The parameterized cross sections are 
defined by equations (\ref{eq:Fnd}) and (\ref{eq:Fdiff}), and parameters 
given in Table \ref{table:gammaParameters}.}
\label{fig:gamma-simvsparam}
\end{figure}

\begin{figure}
\begin{center}
\scalebox{0.65}{\plotone{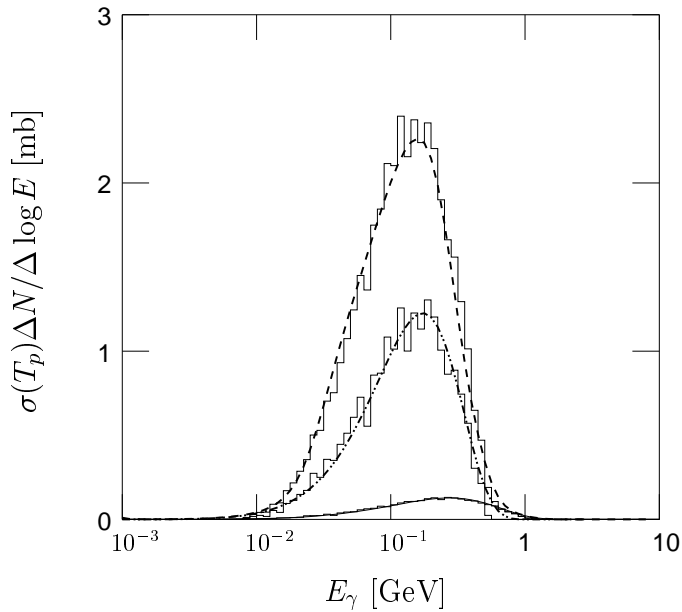}}
\end{center}
\caption{Simulated (histograms) and parameterized 
(lines) gamma-ray inclusive cross sections for 
the resonance-excitation processes at $T_{p}=$ 0.69 GeV.
$\Delta$(1232) (double-dot-dashed line), res(1600) (thin solid line),
and the sum of all model distributions (dashed line). The
parameterized cross sections are defined by equations (\ref{eq:Fdelta})
for both $\Delta$(1232) and res(1600), and parameters are
given in Table \ref{table:gammaParameters}.}
\label{fig:gamma-deltacontrib}
\end{figure}

The simulated gamma-ray inclusive cross sections are superimposed with
the parameterized ones in 
Figures \ref{fig:gamma-simvsparam}a and \ref{fig:gamma-simvsparam}b 
for three mono-energetic protons, for the non-diffractive and
diffractive processes and for the resonance-excitation processes
in Figure \ref{fig:gamma-deltacontrib}.
The agreement is generally good except near the higher and lower kinematical
limits where we find a difference of about 10-20\%. 

\clearpage

\clearpage
\section{Representation of Parameters as Functions of Incident Proton Energy}

The parameterization formulae for secondary particles for 
mono-energetic protons (equations [\ref{eq:Fnd}], [\ref{eq:Fdiff}] and [\ref{eq:Fdelta}]) 
have nine, eight, and five parameters for each $T_p$ for 
non-diffractive, diffractive, and resonance-excitation processes respectively. 
These parameters depend on the
proton kinetic energy, $T_p$. The final step of the parameterization
is to find simple functions representing energy dependence of these 
parameters. Functions obtained by fitting often give values different 
significantly from those found for mono-energetic protons near the
kinematic limits and produce artifacts in the wide range spectral
energy density, as described previously. Some manual adjustments
have been made to control possible artifacts.

\subsection{Parameterized Gamma Ray Spectrum}

\begin{figure}
\begin{center}
\scalebox{0.75}{\plotone{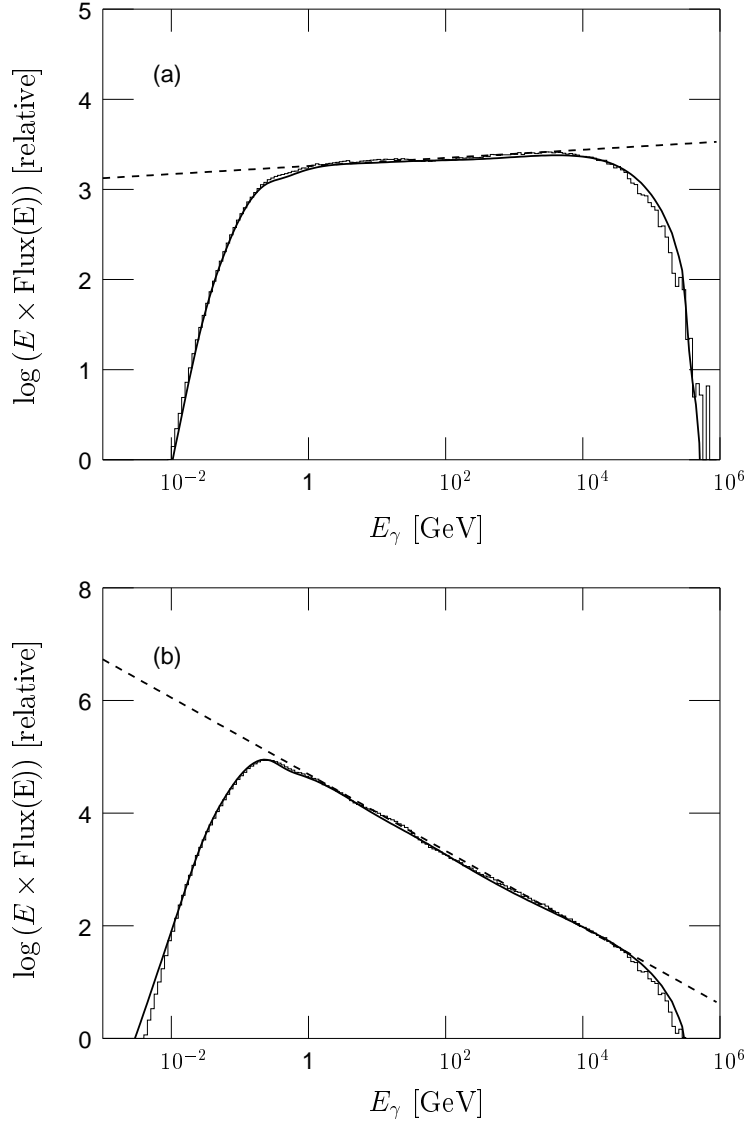}}
\end{center}
\caption{Gamma-ray spectra produced by protons with power-law spectrum 
cutoff at $T_p = 512$ TeV: 
(a) index=2 and (b) index=2.7
by the parametrized model ({\it{solid lines}}) defined by 
equations (\ref{eq:Fnd}), (\ref{eq:Fdiff}), (\ref{eq:Fdelta}), and 
by Table \ref{table:gammaParameters}.
The histograms are those of the simulation by the readjusted model A.
The dashed straight line corresponds to an index of (a) 1.95 and (b) 2.68.}
\label{fig:gamma-spectrum-ind227}
\end{figure}

The final functional representation of inclusive cross sections 
for secondary gamma-rays is given in equations
(\ref{eq:Fnd}), (\ref{eq:Fdiff}), and (\ref{eq:Fdelta}) with
parameters defined as functions of $T_p$ in TeV (not GeV)
in Table \ref{table:gammaParameters}.
The total inclusive gamma-ray spectrum is the sum of the
non-diffractive, diffractive, and resonance-excitation contributions.
The spectrum produced by protons with a continuous spectrum can be 
calculated by summing over the total gamma-ray spectra 
for mono-energetic protons with appropriate spectral weight. 
For example, the spectra for power-law protons extending to 
$T_p = 512$ TeV with index=2 and index=2.7 have been calculated 
and compared with the corresponding histograms produced from the
simulated events in Figure \ref{fig:gamma-spectrum-ind227}. 
The parameterized model reproduces either spectrum within 10\%:
it predicts 10-20\% more gamma-rays than simulation
by the readjusted model A at the higher kinematical limit.
\clearpage
\clearpage
\subsection{Parameterized $e^{\pm}$ and Neutrino Spectra}

The parameterization has been extended to other secondary particles, 
$e^-$, $e^+$, $\nu_{e}$, $\bar{\nu}_{e}$, $\nu_{\mu}$, and $\bar{\nu}_{\mu}$,
in the same way as for gamma-rays.
Their functional formulae are represented by equations (\ref{eq:Fnd}), (\ref{eq:Fdiff}), 
and (\ref{eq:Fdelta}) with the parameters defined in Tables
\ref{table:elecParameters}, \ref{table:posiParameters}, 
\ref{table:nueParameters}, \ref{table:antinueParameters},
\ref{table:numuParameters}, and \ref{table:antinumuParameters},
respectively. We note that no $\pi^{-}$ is produced in 
$\Delta$(1232) decay in readjusted model A and, hence, no $e^{-}$ and
$\bar{\nu}_{e}$ either.  

\begin{figure}
\begin{center}
\scalebox{0.75}{\plotone{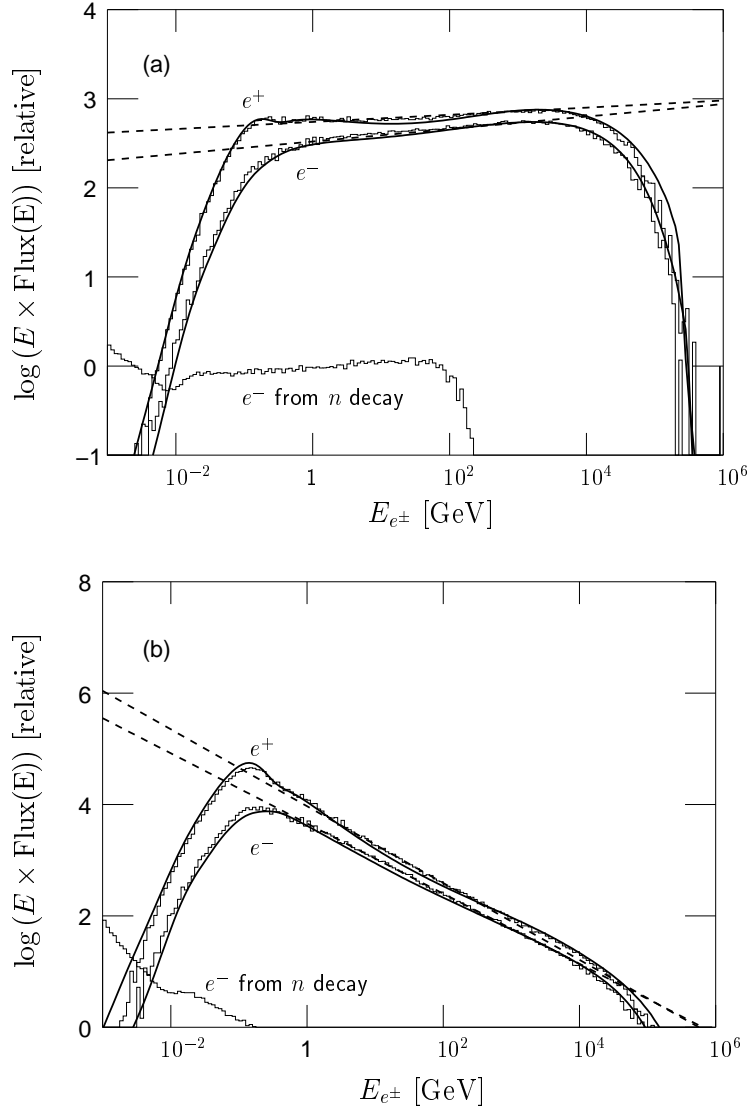}}
\end{center}
\caption{Electron and positron spectra produced by protons with 
power-law spectrum cutoff at $T_p=512$ TeV:
(a) index=2 and (b) index=2.7 by the
readjusted model A and by the parameterized model (solid lines) 
defined by equations (\ref{eq:Fnd}), (\ref{eq:Fdiff}), and (\ref{eq:Fdelta}). 
The parameters are defined in 
Table \ref{table:elecParameters} for electrons and in 
Table \ref{table:posiParameters} for positrons.
The histograms are those of the simulation by the readjusted model A. 
The dashed straight lines correspond to asymptotic 
power-laws: for electrons 1.93 and 2.63 respectively; and
for positrons 1.96 and 2.69 respectively.
The histograms in the lower left corner of (a) and (b) show 
the spectra of 
the electrons from the neutrons produced by the power-law protons.}
\label{fig:elecposi-spectrum-ind227}
\end{figure}

\begin{figure}
\begin{center}
\scalebox{0.75}{\plotone{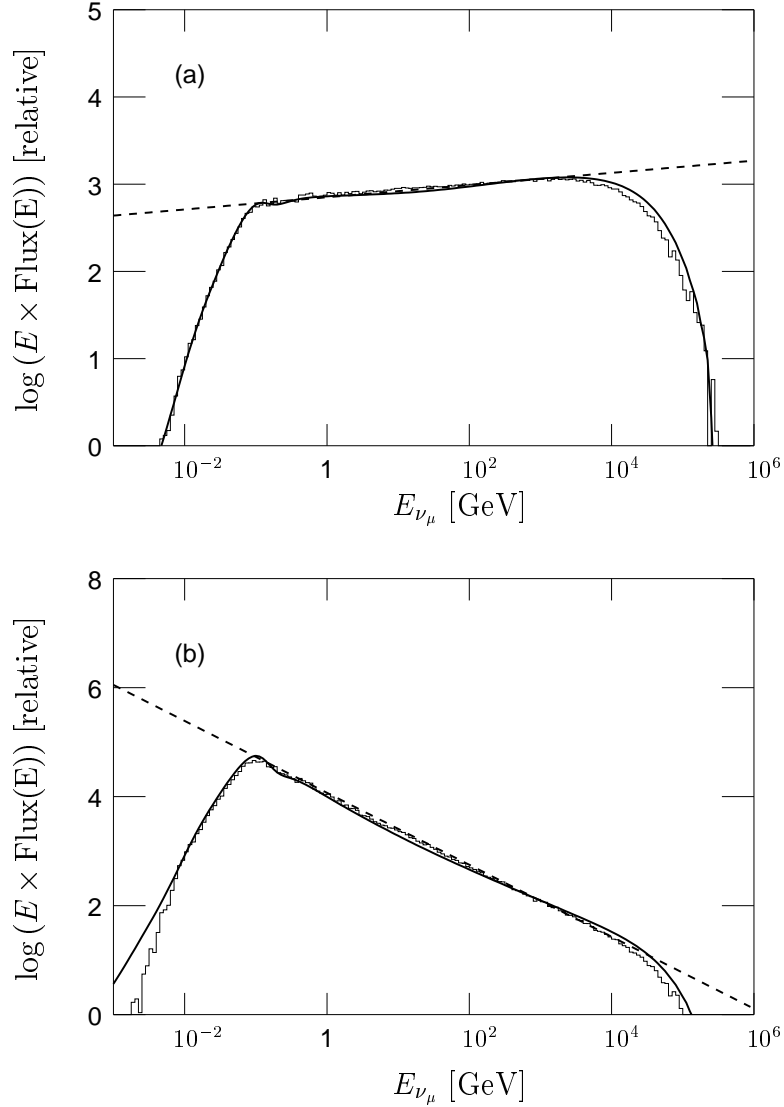}}
\end{center}
\caption{Muon neutrino spectra produced by protons with power-law 
spectrum of (a) index=2 and (b) index=2.7 by the parameterized 
model (solid curves) defined by equations (\ref{eq:Fnd}), (\ref{eq:Fdiff}),
and (\ref{eq:Fdelta}). The parameters are defined in 
Table \ref{table:numuParameters}. The histograms are those of 
the simulation by model A and $\Delta$-excitation. 
The dashed straight lines correspond to index of (a) 1.93 and
(b) 2.66. Note that the muon antineutrino spectra are the same.}
\label{fig:numu-spectrum-ind227}
\end{figure}

We note that the secondary electron and positron spectra from 
charged pion and muon decays have been calculated including
the polarization effect of the weak interaction theory.
The spectra produced by power-law protons of 
index=2.0 and 2.7 ($T_{p}<$ 512 TeV) have been 
computed based on these parameterized models in 
Figure \ref{fig:elecposi-spectrum-ind227} for $e^{\pm}$ 
and Figure \ref{fig:numu-spectrum-ind227} for $\nu_{\mu}$. 
We note in Figure \ref{fig:elecposi-spectrum-ind227} that more $e^{+}$
are produced than $e^{-}$ throughout their spectra. This is
largely due to the charge conservation and enhanced by the fact that 
we have neglected $\alpha$-particles and neutron decays. The number 
of electrons produced in the $p-p$ interaction will match that of 
positrons if we include electrons coming out of neutron decays. 

For given $T_{p}$, electrons from neutron decays
have low energy (mostly with $E<10$~MeV) as shown by the lower 
histograms in Figure \ref{fig:elecposi-spectrum-ind227}.
They do not contribute to the high energy gamma-ray spectrum.
\clearpage

\begin{figure}
\begin{center}
\scalebox{0.75}{\plotone{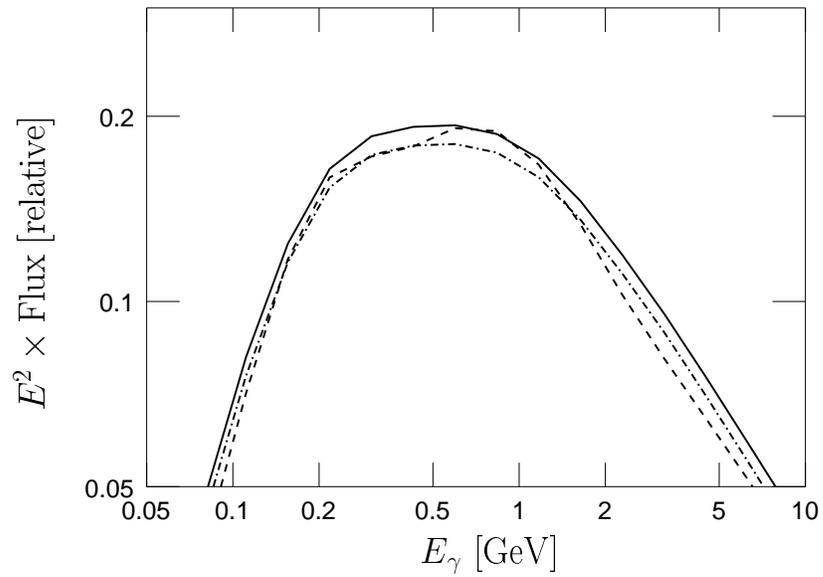}}
\end{center}
\caption{The $\pi^0$ gamma-ray spectra in the Galactic ridge
obtained with Galprop with the parameter set, galdef 500180:
the Galprop built-in $p-p$ interaction model (dashed line);
the present parameterized $p-p$ interaction model (dash-dot line);
the present parameterized model 
normalized to the dashed line in the energy band 0.7-0.8 GeV (solid line).}
\label{fig:piondecayskymap}
\end{figure}
\clearpage
\section{Application to Galactic Diffuse Gamma-Ray Emission}

We have replaced the $\pi^0$ production subroutine of Galprop
\citep{Galprop1, Galprop2} with the present parameterized model 
and compared the Galactic diffuse gamma-ray spectra of $\pi^0$
origin with that by the built-in subroutine. A common parameter
set, galdef 500180 described in \cite{SMR04}, has been used in
the two calculations.  

As shown in Figure \ref{fig:piondecayskymap}, the present
parameterized model gives a flatter and smoother spectral energy
distribution between $E_{\gamma}=0.3-2$ GeV, smaller gamma-ray
yield between $0.5-1.3$ GeV, and a higher power-law index between
$1-5$ GeV than the Galprop built-in model. In \citet{Kamae05},
gamma-ray spectrum of the Galprop built-in model was compared with
that of model A, after being normalized in the energy range $E<300$ MeV.
This normalization enhanced gamma-ray yield in the GeV range 
relative to sub-GeV range. The Galprop built-in 
$p-p$ interaction model has been tuned to reproduce accelerator 
experiments better than model A of \citet{Kamae05} near the threshold. 
We have included the resonance contributions in the present model 
to improve this shortcoming near the threshold. 
Hence we have change the normalization point to the peak region
in $E^2 d{\rm{flux}}/dE$, $E_{\gamma}=0.7-0.8$ GeV, where 
gamma-rays from $\pi^{0}$ decays are expected to dominate 
(see Figure \ref{fig:piondecayskymap}, dashed curve).
The present model thus normalized gives $\sim 20$\% higher gamma-ray 
yield than Galprop at 2 GeV: this is substantially lower than
the difference (about 50\%) shown in Figure 7 of \citet{Kamae05}.
The gamma-ray power-law index of the present model is harder by about
0.05 than that by the Galprop built-in model, just as the model A of
\citet{Kamae05} predicts.

\section{Conclusion and Future Prospects}

We have presented the inclusive cross sections of stable secondary 
particles ($\gamma$, $e^\pm$, $\nu_e$, $\bar{\nu}_e$, $\nu_\mu$,
and $\bar{\nu}_\mu$) produced by the $p-p$ interaction in
parameterized formulae. They facilitates computation of secondary
particle spectra for arbitrary proton spectra as shown for Galactic
diffuse gamma-ray spectrum (Figure \ref{fig:piondecayskymap}).
Various effects that these secondary particles may have in astronomical 
environments can also be calculated at a higher precision. The
formulae incorporate all important known features of the $p-p$
interaction up to about $T_p = 500$ TeV and hence will also be useful
in calculating background when searching for new phenomena.

The parameterized model predicts all secondary particle spectra 
to have harder power-law indices than that of the incident proton 
and their inclusive cross sections to be larger than those expected 
from the old $p-p$ interaction models. When used to replace the $p-p$
subroutine in Galprop \citep{Galprop1, Galprop2}, the model gives a
flatter spectral energy density distribution between $0.3-5$ GeV.
The absolute gamma-ray yield predicted by the model is smaller 
than that by the Galprop model for $E_\gamma <1.5$~GeV but higher for 
$E_\gamma >1.5$~GeV. If normalized near the peak  
in $E^2 d{\rm{flux}}/dE$ ($E_{\gamma}=0.7-0.8$~GeV),
the parameterized model gives $\sim 20$\% higher gamma-ray yield 
at 2~GeV than the model of Galprop: our model with this normalization
can account for $\sim 20$\%, not $\sim 50$\% as was claimed
in \citet{Kamae05}, of the discrepancy between the diffuse 
Galactic Ridge gamma-ray spectrum observed by EGRET 
and model predictions for the proton spectrum near the solar 
system (the power-law index $\sim 2.7$) \citep{Hunter97}.
We note the discrepancy will be reduced 
by inclusion of the inverse Compton component (see, for example,
a model by \citet{SMR04}).

The present model also predicts more $e^{+}$ than $e^{-}$ at
energies higher wherever those produced by $p-p$ interaction become
comparable in flux to primary $e^{-}$ and $e^{+}$. 
The formulae and parameters given in the appendices are available
as supplementary on-line material both in C language format
and as C subroutine.
We are currently parametrizing the angular distribution of gamma-rays
relative to the incident proton direction. The results will be published 
elsewhere.

\section{Acknowledgments}

The authors would like to acknowledge valuable discussions with
and comments received from M. Asai, E. Bloom, P. Carlson, J. Chiang, 
J. Cohen-Tanugi, S. Digel, E. do Couto e Silva, I. Grenier, G. Madejski, 
I. Moskalenko, P. Nolan, O. Reimer, D. Smith, F. Stecker, 
A. Strong, H. Tajima, and L. Wai.  They wish to thank the anonimous
referee for valuable comments and suggestions.

\clearpage
\begin{table}
\caption{Constants for equations \ref{eq:sigmaNonDiff}, \ref{eq:sigmaTotDiff}, \ref{eq:sigmaDelta}
and \ref{eq:sigmaReson}.}
\scalebox{0.92}{\begin{tabular}{lllllll}
\hline\hline
$a_{0}=0.1176$ & $b_{0}=11.34$ & $c_{0}=28.5$ & $d_{0}=0.3522$ & $e_{0}=5.922$ & $f_{0}=0.0834$ & $g_{0}=0.0004257$ \\
$a_{1}=0.3829$ & $b_{1}=23.72$ & $c_{1}=-6.133$ & $d_{0}=0.1530$ & $e_{1}=1.632$ & $f_{1}=9.5$ & $g_{1}=4.5$ \\
$a_{2}=23.10$ & & $c_{2}=1.464$ & $d_{2}=1.498$ & & $f_{2}=-5.5$ & $g_{2}=-7.0$ \\
$a_{3}=6.454$ & & & $d_{3}=2.0$ & & $f_{3}=1.68$ & $g_{3}=2.1$ \\
$a_{4}=-5.764$ & & & $d_{4}=30.0$ & & $f_{4}=3134$ & $g_{4}=503.5$ \\
$a_{5}=-23.63$ & & & $d_{5}=3.155$ \\
$a_{6}=94.75$ & & & $d_{6}=1.042$ \\
$a_{7}=0.02667$ \\
\hline\hline
\end{tabular}}
\label{table:sigmaPPconstants}
\end{table}

\begin{table}
\caption{Kinematic limit parameters for the non-diffractive process. 
Proton kinetic energy $T_{p}$ is in GeV.}
\label{table:E_max-Delta}
\begin{tabular}{cccc}
\hline\hline
Particle & $L_{\rm{max}}$ & $W_{\rm{ND,l}}$ & $W_{\rm{ND,h}}$ \\
\hline
$\gamma$ & 0.96$\log_{10}(T_{p})$ & 15 & 44 \\
$e^{-}$ & 0.96$\log_{10}(T_{p})$ & 20 & 45 \\
$e^{+}$ & 0.94$\log_{10}(T_{p})$ & 15 & 47 \\
$\nu_{e}$ & 0.98$\log_{10}(T_{p})$ & 15 & 42 \\
$\bar{\nu}_{e}$ & 0.98$\log_{10}(T_{p})$ & 15 & 40 \\
$\nu_{\mu}$ & 0.94$\log_{10}(T_{p})$ & 20 & 45 \\
$\bar{\nu}_{\mu}$ & 0.98$\log_{10}(T_{p})$ & 15 & 40 \\
\hline
\end{tabular}
\end{table}

\begin{table}
\caption{Parameters describing gamma-ray spectra for arbitrary proton energy.}
\label{table:gammaParameters}
\scalebox{0.60}{\begin{tabular}{ll}
\hline\hline
Parameter & Formulae as functions of the proton kinetic energy 
($y=\log_{10}(T_{p})$) in TeV. \\
\hline
\multicolumn{2}{l}{Non-diff., equation \ref{eq:Fnd}} \\
\hline
$a_0$ & $-0.51187(y + 3.3) + 7.6179(y + 3.3)^{2} - 2.1332(y + 3.3)^{3} + 0.22184(y + 3.3)^{4}$ \\
$a_1$ & $-1.2592\tpow{-5} + 1.4439\tpow{-5}\exp{(-0.29360(y + 3.4))} + 5.9363\tpow{-5}/(y + 4.1485) + 2.2640\tpow{-6}y - 3.3723\tpow{-7}y^{2}$\\
$a_2$ & $-174.83 + 152.78\log_{10}{(1.5682(y + 3.4))} - 808.74/(y + 4.6157)$ \\
$a_3$ & $0.81177 + 0.56385y + 0.0040031y^{2} - 0.0057658y^{3} + 0.00012057y^{4}$ \\
$a_4$ & $0.68631(y + 3.32) + 10.145(y + 3.32)^{2} - 4.6176(y + 3.32)^{3} + 0.86824(y + 3.32)^{4} - 0.053741(y + 3.32)^{5}$ \\
$a_5$ & $9.0466\tpow{-7} + 1.4539\tpow{-6}\log_{10}{(0.015204(y + 3.4))} + 1.3253\tpow{-4}/(y + 4.7171)^{2} - 4.1228\tpow{-7}y + 2.2036\tpow{-7}y^{2}$ \\
$a_6$ & $-339.45 + 618.73\log_{10}{(0.31595(y + 3.9))} + 250.20/(y + 4.4395)^{2}$ \\
$a_7$ & $-35.105 + 36.167y - 9.3575y^{2} + 0.33717y^{3}$ \\
$a_8$ & $0.17554 + 0.37300y - 0.014938y^{2} + 0.0032314y^{3} + 0.0025579y^{4}$ \\
\hline
$r(y)$ & $3.05\exp{(-107((y + 3.25)/(1 + 8.08(y + 3.25)))^{2})}$ for $T_{p}\leq 1.95$ GeV \\
       & 1.01 for $T_{p} > 1.95$ GeV \\
\hline
\multicolumn{2}{l}{Diffraction, equation \ref{eq:Fdiff}} \\
\hline 
$b_0$ & $60.142\tanh{(-0.37555(y + 2.2))} - 5.9564(y + 0.59913)^{2} + 6.0162\tpow{-3}(y + 9.4773)^{4}$ \\
$b_1$ & $35.322 + 3.8026\tanh{(-2.5979(y + 1.9))} - 2.1870\tpow{-4}(y + 369.13)^{2}$ \\
$b_2$ & $-15.732 - 0.082064\tanh{(-1.9621(y + 2.1))} + 2.3355\tpow{-4}(y + 252.43)^{2}$ \\
$b_3$ & $-0.086827 + 0.37646\exp{(-0.53053((y + 1.0444)/(1.0 + 0.27437(y + 1.0444)))^2)}$ \\
      & Important: $b_{0},\ldots,b_{3}$ = 0 for $T_{p} < 5.52$ GeV \\
$b_4$ & $2.5982 + 0.39131(y + 2.95)^{2} - 0.0049693(y + 2.95)^{4} + 0.94131\exp{(-24.347(y + 2.45 - 0.19717(y + 2.45)^{2})^{2})}$ \\
$b_5$ & $0.11198 - 0.64582y + 0.16114y^{2} + 2.2853\exp{(-0.0032432((y - 0.83562)/(1.0 + 0.33933(y - 0.83562)))^{2})}$ \\
$b_6$ & $1.7843 + 0.91914y + 0.050118y^{2} + 0.038096y^{3} - 0.027334y^{4} - 0.0035556y^{5} + 0.0025742y^{6}$ \\
$b_7$ & $-0.19870 - 0.071003y + 0.019328y^{2} - 0.28321\exp{(-6.0516(y + 1.8441)^{2})}$ \\
\hline
\multicolumn{2}{l}{$\Delta$(1232), equation \ref{eq:Fdelta}} \\
\hline
$c_0$ & $2.4316\exp{(-69.484((y + 3.1301)/(1.0 + 1.24921(y + 3.1301)))^{2})} - 6.3003 - 9.5349/y + 0.38121y^{2}$ \\
$c_1$ & $56.872 + 40.627y + 7.7528y^{2}$ \\
$c_2$ & $-5.4918 - 6.7872\tanh{(4.7128(y + 2.1))} + 0.68048y$ \\
$c_3$ & $-0.36414 + 0.039777y$ \\
$c_4$ & $-0.72807 - 0.48828y - 0.092876y^{2}$ \\
\hline
\multicolumn{2}{l}{Res1600), equation \ref{eq:Fdelta}} \\
\hline
$d_0$ & $3.2433\exp{(-57.133((y + 2.9507)/(1.0 + 1.2912(y + 2.9507)))^{2})} - 1.0640 - 0.43925y$ \\
$d_1$ & $16.901 + 5.9539y - 2.1257y^{2} - 0.92057y^{3}$ \\
$d_2$ & $-6.6638 - 7.5010\tanh{(30.322(y + 2.1))} + 0.54662y$ \\
$d_3$ & $-1.50648 - 0.87211y - 0.17097y^{2}$ \\
$d_4$ & $0.42795 + 0.55136y + 0.20707y^{2} + 0.027552y^{3}$ \\
\hline\hline
\end{tabular}}
\end{table}

\begin{table}
\caption{Parameter describing electron spectra for arbitrary proton energy.}
\label{table:elecParameters}
\scalebox{0.65}{\begin{tabular}{ll}
\hline\hline
Parameters & Formulae as functions of the proton kinetic energy 
($y=\log_{10}(T_{p})$) in TeV. \\
\hline
\multicolumn{2}{l}{Non-diff., equation \ref{eq:Fnd}} \\
\hline
$a_0$ & $-0.018639(y + 3.3) + 2.4315(y + 3.3)^{2} - 0.57719(y + 3.3)^{3} + 0.063435(y + 3.3)^{4}$ \\
$a_1$ & $7.1827\tpow{-6} - 3.5067\tpow{-6}y + 1.3264\tpow{-6}y^{2} - 3.3481\tpow{-7}y^{3} + 2.3551\tpow{-8}y^{4} + 3.4297\tpow{-8}y^{5}$ \\
$a_2$ & $563.91 - 362.18\log_{10}{(2.7187(y + 3.4))} - 2.8924\tpow{4}/(y + 7.9031)^{2}$ \\
$a_3$ & $0.52684 + 0.57717y + 0.0045336y^{2} - 0.0089066y^{3}$ \\
$a_4$ & $0.36108(y + 3.32) + 1.6963(y + 3.32)^{2} - 0.074456(y + 3.32)^{3} - 0.071455(y + 3.32)^{4} + 0.010473(y + 3.32)^{5}$ \\
$a_5$ & $9.7387\tpow{-5} + 7.8573\tpow{-5}\log_{10}{(0.0036055(y + 4.3))} + 0.00024660/(y + 4.9390) - 3.8097\tpow{-7}y^{2}$ \\
$a_6$ & $-273.00 - 106.22\log_{10}{(0.34100(y + 3.4))} + 89.037y - 12.546y^{2}$ \\
$a_7$ & $432.53 - 883.99\log_{10}{(0.19737(y + 3.9))} - 4.1938\tpow{4}/(y + 8.5518)^2$ \\
$a_8$ & $-0.12756 + 0.43478y - 0.0027797y^{2} - 0.0083074y^{3}$ \\
\hline
$r(y)$ & $3.63\exp{(-106((y + 3.26)/(1 + 9.21(y + 3.26)))^{2})} - 0.182y - 0.175y^{2}$ for $T_{p}\leq 15.6$ GeV \\
       & 1.01 for $T_{p} > 15.6$ GeV \\
\hline
\multicolumn{2}{l}{Diffraction, equation \ref{eq:Fdiff}} \\
\hline 
$b_0$ & $0.20463\tanh{(-6.2370(y + 2.2))} - 0.16362(y + 1.6878)^{2} + 3.5183\tpow{-4}(y + 9.6400)^{4}$ \\
$b_1$ & $1.6537 + 3.8530\exp{(-3.2027((y + 2.0154)/(1.0 + 0.62779(y + 2.0154)))^{2})}$ \\
$b_2$ & $-10.722 - 0.082672\tanh{(-1.8879(y + 2.1))} + 1.4895\tpow{-4}(y + 256.63)^{2}$ \\
$b_3$ & $-0.023752 - 0.51734\exp{(-3.3087((y + 1.9877)/(1.0 + 0.40300(y + 1.9877)))^2)}$ \\
      & Important: $b_{0},\ldots,b_{3}$ = 0 for $T_{p} < 5.52$ GeV \\
$b_4$ & $0.94921 + 0.12280(y + 2.9)^{2} - 7.1585\tpow{-4}(y + 2.9)^{4} + 0.52130\log_{10}{(y + 2.9)}$ \\
$b_5$ & $-4.2295 - 1.0025\tanh{(9.0733(y + 1.9))} - 0.11452*(y - 62.382)$ \\
$b_6$ & $1.4862 + 0.99544y - 0.042763y^{2} - 0.0040065y^{3} + 0.0057987y^{4}$ \\
$b_7$ & $6.2629 + 6.9517\tanh{(-0.36480(y + 2.1))} - 0.026033*(y - 2.8542)$ \\
\hline
\multicolumn{2}{l}{Res1600), equation \ref{eq:Fdelta}} \\
\hline
$d_0$ & $0.37790\exp{(-56.826((y + 2.9537)/(1.0 + 1.5221(y + 2.9537)))^{2})} - 0.059458 + 0.0096583y^{2}$ \\
$d_1$ & $-5.5135 - 3.3988y$ \\
$d_2$ & $-7.1209 - 7.1850\tanh{(30.801(y + 2.1))} + 0.35108y$ \\
$d_3$ & $-6.7841 - 4.8385y - 0.91523y^{2}$ \\
$d_4$ & $-134.03 - 139.63y - 48.316y^{2} - 5.5526y^{3}$ \\
\hline
\end{tabular}}
\end{table}

\begin{table}
\caption{Parameters describing positron spectra for arbitrary proton energy.}
\label{table:posiParameters}
\scalebox{0.60}{\begin{tabular}{ll}
\hline\hline
Parameters & Formulae as functions of the proton kinetic energy 
($y=\log_{10}(T_{p})$) in TeV. \\
\hline
\multicolumn{2}{l}{Non-diff., equation \ref{eq:Fnd}} \\
\hline
$a_0$ & $-0.79606(y + 3.3) + 7.7496(y + 3.3)^{2} - 3.9326(y + 3.3)^{3} + 0.80202(y + 3.3)^{4} - 0.054994(y + 3.3)^{5}$ \\
$a_1$ & $6.7943\tpow{-6} - 3.5345\tpow{-6}y + 6.0927\tpow{-7}y^{2} + 2.0219\tpow{-7}y^{3} + 5.1005\tpow{-8}y^{4} - 4.2622\tpow{-8}y^{5}$ \\
$a_2$ & $44.827 - 81.378\log_{10}{(0.027733(y + 3.5))} - 1.3886\tpow{4}/(y + 8.4417)$ \\
$a_3$ & $0.52010 + 0.59336y + 0.012032y^{2} - 0.0064242y^{3}$ \\
$a_4$ & $2.1361(y + 3.32) + 1.8514(y + 3.32)^{2} - 0.47872(y + 3.32)^{3} + 0.0032043(y + 3.32)^{4} + 0.0082955(y + 3.32)^{5}$ \\
$a_5$ & $1.0845\tpow{-6} + 1.4336\tpow{-6}\log_{10}{(0.0077255(y + 4.3))} + 1.3018\tpow{-4}/(y + 4.8188)^{2} + 9.3601\tpow{-8}y$ \\
$a_6$ & $-267.74 + 14.175\log_{10}{(0.35391(y + 3.4))} + 64.669/(y - 7.7036)^{2}$ \\
$a_7$ & $138.26 - 539.84\log_{10}{(0.12467(y + 3.9))} - 1.9869\tpow{4}/(y + 7.6884)^2 + 1.0675y^{2}$ \\
$a_8$ & $-0.14707 + 0.40135y + 0.0039899y^{2} - 0.0016602y^{3}$ \\
\hline
$r(y)$ & $2.22\exp{(-98.9((y + 3.25)/(1 + 10.4(y + 3.25)))^{2})}$ for $T_{p}\leq 5.52$ GeV \\
           & 1.0 for $T_{p} > 5.52$ GeV \\
\hline
\multicolumn{2}{l}{Diffraction, equation \ref{eq:Fdiff}} \\
\hline 
$b_0$ & $29.192\tanh{(-0.37879(y + 2.2))} - 3.2196(y + 0.67500)^{2} + 0.0036687(y + 9.0824)^{4}$ \\
$b_1$ & $-142.97 + 147.86\exp{(-0.37194((y + 1.8781)/(1.0 + 3.8389(y + 1.8781)))^{2})}$ \\
$b_2$ & $-14.487 - 4.2223\tanh{(-13.546(y + 2.2))} + 1.6988\tpow{-4}(y + 234.65)^{2}$ \\
$b_3$ & $-0.0036974 - 0.41976\exp{(-6.1527((y + 1.8194)/(1.0 + 0.99946(y + 1.8194)))^2)}$ \\
      & Important: $b_{0},\ldots,b_{3}$ = 0 for $T_{p} < 11.05$ GeV \\
$b_4$ & $1.8108 + 0.18545(y + 2.9)^{2} - 0.0020049(y + 2.9)^{4} + 0.85084\exp{(-14.987(x + 2.29 - 0.18967(x + 2.29)^{2})^{2})}$ \\
$b_5$ & $2.0404 - 0.51548\tanh{(2.2758(y + 1.9))} - 0.035009/(y - 6.6555)$ \\
$b_6$ & $1.5258 + 1.0132y - 0.064388y^{2} - 0.0040209y^{3} - 0.0082772y^{4}$ \\
$b_7$ & $3.0551 + 3.5240\tanh{(-0.36739(y + 2.1))} - 0.13382*(y - 2.7718)$ \\
\hline
\multicolumn{2}{l}{$\Delta$(1232), equation \ref{eq:Fdelta}} \\
\hline
$c_0$ & $2.9841\exp{(-67.857((y + 3.1272)/(1.0 + 0.22831(y + 3.1272)))^{2})} - 6.5855 - 9.6984/y + 0.41256y^{2}$ \\
$c_1$ & $6.8276 + 5.2236y + 1.4630y^{2}$ \\
$c_2$ & $-6.0291 - 6.4581\tanh(5.0830(y + 2.1)) + 0.46352y$ \\
$c_3$ & $0.59300 + 0.36093y$ \\
$c_4$ & $0.77368 + 0.44776y + 0.056409y^{2}$ \\
\hline
\multicolumn{2}{l}{Res1600), equation \ref{eq:Fdelta}} \\
\hline
$d_0$ & $1.9186\exp{(-56.544((y + 2.9485)/(1.0 + 1.2892(y + 2.9485)))^{2})} - 0.23720 + 0.041315y^{2}$ \\
$d_1$ & $-4.9866 - 3.1435y$ \\
$d_2$ & $-7.0550 - 7.2165\tanh(31.033(y + 2.1)) + 0.38541y$ \\
$d_3$ & $-2.8915 - 2.1495y - 0.45006y^{2}$ \\
$d_4$ & $-1.2970 - 0.13947y - 0.41197y^{2} - 0.10641y^{3}$ \\
\hline\hline
\end{tabular}}
\end{table}

\begin{table}
\caption{Parameters describing electron neutrino spectra for arbitrary proton energy.}
\label{table:nueParameters}
\scalebox{0.65}{\begin{tabular}{ll}
\hline\hline
Parameters & Formulae as functions of the proton kinetic energy 
($y=\log_{10}(T_{p})$) in TeV. \\
\hline
\multicolumn{2}{l}{Non-diff., equation \ref{eq:Fnd}} \\
\hline
$a_0$ & $0.0074087 + 2.9161(y + 3.31) + 0.99061(y + 3.31)^{2} - 0.28694(y + 3.31)^{3} + 0.038799(y + 3.31)^{4}$ \\
$a_1$ & $-3.2480\tpow{-5} + 7.1944\tpow{-5}\exp{(-0.21814(y + 3.4))} + 2.0467\tpow{-5}/(y + 4.1640) + 5.6954\tpow{-6}y - 3.4105\tpow{-7}y^{2}$ \\
$a_2$ & $-230.50 + 58.802y - 9.9393y^{2} + 1.2473y^{3} - 0.26322y^{4}$ \\
$a_3$ & $0.45064 + 0.56930y + 0.012428y^{2} - 0.0070889y^{3}$ \\
$a_4$ & $-0.011883 + 1.7992(y + 3.32) + 3.5264(y + 3.32)^{2} - 1.7478(y + 3.32)^{3} + 0.32077(y + 3.32)^{4} - 0.017667(y + 3.32)^{5}$ \\
$a_5$ & $-1.6238\tpow{-7} + 1.8116\tpow{-6}\exp{(-0.30111(y + 3.4))} + 9.6112\tpow{-5}/(y + 4.8229)^{2}$ \\
$a_6$ & $-261.30 - 43.351\log_{10}{(0.35298(y + 3.4))} + 70.925/(y - 8.7147)^{2}$ \\
$a_7$ & $184.45 - 1473.6/(y + 6.8788) - 4.0536y^{2}$ \\
$a_8$ & $-0.24019 + 0.38504y + 0.0096869y^{2} - 0.0015046y^{3}$ \\
\hline
$r(y)$ & $0.329\exp{(-247((y + 3.26)/(1 + 6.56(y + 3.26)))^{2})} - 0.957y - 0.229y^{2}$ for $T_{p}\leq 7.81$ GeV \\
      & 1.0 for $T_{p} > 7.81$ GeV \\
\hline
\multicolumn{2}{l}{Diffraction, equation \ref{eq:Fdiff}} \\
\hline 
$b_0$ & $53.809\tanh{(-0.41421(y + 2.2))} - 6.7538(y + 0.76010)^{2} + 0.0088080(y + 8.5075)^{4}$ \\
$b_1$ & $-50.211 + 55.131\exp{(1.3651((y + 1.8901)/(1.0 + 4.4440(y + 1.8901)))^{2})}$ \\
$b_2$ & $-17.231 + 0.041100\tanh{(7.9638(y + 1.9))} - 0.055449y + 2.5866\tpow{-4}(y + 250.68)^{2}$ \\
$b_3$ & $12.335 - 12.893\exp{(-1.4412((y + 1.8998)/(1.0 + 5.5969(y + 1.8998)))^2)}$ \\
      & Important: $b_{0},\ldots,b_{3}$ = 0 for $T_{p} < 11.05$ GeV \\
$b_4$ & $1.3558 + 0.46601(y + 2.95) + 0.052978(y + 2.2)^{2} + 0.79575\exp{(-5.4007(y + 2.2 + 4.6121(x + 2.2)^{2})^{2})}$ \\
$b_5$ & $1.8756 - 0.42169\tanh{(1.6100(y + 1.9))} - 0.051026*(y - 3.9573)$ \\
$b_6$ & $1.5016 + 1.0118y - 0.072787y^{2} - 0.0038858y^{3} + 0.0093650y^{4}$ \\
$b_7$ & $4.9735 + 5.5674\tanh{(-0.36249(y + 2.1))} - 0.20660*(y - 2.8604)$ \\
\hline
\multicolumn{2}{l}{$\Delta$(1232), equation \ref{eq:Fdelta}} \\
\hline
$c_0$ & $2.8290\exp{(-71.339((y + 3.1282)/(1.0 + 0.48420(y + 3.1282)))^{2})} - 9.6339 - 15.733/y + 0.52413y^{2}$ \\
$c_1$ & $-24.571 - 15.831y - 2.1200y^{2}$ \\
$c_2$ & $-5.9593 - 6.4695\tanh{(4.7225(y + 2.1))} + 0.50003y$ \\
$c_3$ & $0.26022 + 0.24545y$ \\
$c_4$ & $0.076498 + 0.061678y + 0.0040028y^{2}$ \\
\hline
\multicolumn{2}{l}{Res1600), equation \ref{eq:Fdelta}} \\
\hline
$d_0$ & $1.7951\exp{(-57.260((y + 2.9509)/(1.0 + 1.4101(y + 2.9509)))^{2})} - 0.58604 - 0.23868y$ \\
$d_1$ & $-2.6395 - 1.5105y + 0.22174y^{2}$ \\
$d_2$ & $-7.0512 - 7.1970\tanh{(31.074(y + 2.1))} + 0.39007y$ \\
$d_3$ & $-1.4271 - 1.0399y - 0.24179y^{2}$ \\
$d_4$ & $0.74875 + 0.63616y + 0.17396y^{2} + 0.017636y^{3}$ \\
\hline\hline
\end{tabular}}
\end{table}

\begin{table}
\caption{Parameters describing electron anti-neutrino spectra for arbitrary proton energy.}
\label{table:antinueParameters}
\scalebox{0.65}{\begin{tabular}{ll}
\hline\hline
Parameters & Formulae as functions of the proton kinetic energy 
($y=\log_{10}(T_{p})$) in TeV. \\
\hline
\multicolumn{2}{l}{Non-diff., equation \ref{eq:Fnd}} \\
\hline
$a_0$ & $0.0013113 + 0.36538(y + 3.31) + 1.5178(y + 3.31)^{2} - 0.20668(y + 3.31)^{3} + 0.024255(y + 3.31)^{4}$ \\
$a_1$ & $-4.7833\tpow{-6} + 4.5837\tpow{-5}\exp{(-0.42980(y + 3.4))} + 6.1559\tpow{-6}/(y + 4.1731) + 1.1928\tpow{-6}y$ \\
$a_2$ & $-245.22 + 73.223y - 19.652y^{2} + 0.083138y^{3} + 0.71561y^{4} $ \\
$a_3$ & $0.45232 + 0.52934y + 0.010078y^{2} - 0.0017092y^{3}$ \\
$a_4$ & $-0.0025734 + 0.38424(y + 3.32) + 1.5517(y + 3.32)^{2} + 0.17336(y + 3.32)^{3} - 0.17160(y + 3.32)^{4} + 0.021059(y + 3.32)^{5}$ \\
$a_5$ & $4.7673\tpow{-5} + 5.4936\tpow{-5}\log{(0.0067905(y + 4.3))} + 0.00020740/(y + 4.9772)$ \\
$a_6$ & $-270.30 - 114.47\log_{10}{(0.34352(y + 3.4))} + 80.085y - 7.9240y^{2}$ \\
$a_7$ & $3272.9 - 2.9161\tpow{5}/(y + 87.847) - 6.2330y^{2}$ \\
$a_8$ & $-0.17787 + 0.36771y - 0.025397y^{2} + 0.0019238y^{3} + 0.0032725y^{4}$ \\
\hline
$r(y)$ & $2.67\exp{(-45.7((y + 3.27)/(1 + 6.59(y + 3.27)))^{2})} - 0.301y - 0.208y^{2}$ for $T_{p}\leq 15.6$ GeV \\
           & 1.0 for $T_{p} > 15.6$ GeV \\
\hline
\multicolumn{2}{l}{Diffraction, equation \ref{eq:Fdiff}} \\
\hline 
$b_0$ & $41.307\tanh{(-0.37411(y + 2.2))} - 4.1223(y + 0.55505)^{2} + 0.0042652(y + 9.2685)^{4}$ \\
$b_1$ & $-132.50 + 142.12\exp{(-8.0289((y + 1.9196)/(1.0 + 11.530(y + 1.9196)))^{2})}$ \\
$b_2$ & $-17.223 + 0.011285\tanh{(-69.746(y + 1.9))} - 0.048233y + 2.5881\tpow{-4}(y + 250.77)^{2}$ \\
$b_3$ & $8.1991 - 9.6437\exp{(-45.261((y + 1.9292)/(1.0 + 16.682(y + 1.9292)))^2)}$ \\
      & Important: $b_{0},\ldots,b_{3}$ = 0 for $T_{p} < 11.05$ GeV \\
$b_4$ & $0.55919 + 0.36647(y + 2.95)^{2} + 0.056194(y + 2.95)^{4} + 0.49957\exp{(-5.5317(y + 2.2 + 0.43867(y + 2.2)^{2})^{2})}$ \\
$b_5$ & $1.2544 - 0.52362\tanh{(2.7638(y + 1.9))} + 0.055837*(y - 17.638)$ \\
$b_6$ & $1.4788 + 1.0278y - 0.092852y^{2} - 0.0062734y^{3} + 0.011920y^{4}$ \\
$b_7$ & $5.1651 + 5.7398\tanh{(-0.37356(y + 2.1))} - 0.22234*(y - 2.7889)$ \\
\hline
\multicolumn{2}{l}{Res1600), equation \ref{eq:Fdelta}} \\
\hline
$d_0$ & $0.36459\exp{(-58.210((y + 2.9537)/(1.0 + 1.4320(y + 2.9537)))^{2})} - 0.11283 - 0.046244y$ \\
$d_1$ & $-9.5066 - 5.4655y - 0.31769y^{2}$ \\
$d_2$ & $-7.1831 - 7.1551tanh(30.354(y + 2.1)) + 0.33757y$ \\
$d_3$ & $2.7938 + 1.6992y + 0.20161y$ \\
$d_4$ & $0.61878 + 0.62371y + 0.18913y^{2} + 0.019118y^{3}$ \\
\hline\hline
\end{tabular}}
\end{table}

\begin{table}
\caption{Parameters describing muon neutrino spectra for arbitrary proton energy.}
\label{table:numuParameters}
\scalebox{0.65}{\begin{tabular}{ll}
\hline\hline
Parameters & Formulae as functions of the proton kinetic energy 
($y=\log_{10}(T_{p})$) in TeV. \\
\hline
\multicolumn{2}{l}{Non-diff., equation \ref{eq:Fnd}} \\
\hline
$a_0$ & $-0.63611(y + 3.3) + 9.9015(y + 3.3)^{2} - 4.5897(y + 3.3)^{3} + 0.91778(y + 3.3)^{4} - 0.060724(y + 3.3)^{4}$ \\
$a_1$ & $6.8700\tpow{-6} - 2.8245\tpow{-6}y + 7.6032\tpow{-7}y^{2} - 3.2953\tpow{-7}y^{3} + 7.4292\tpow{-8}y^{4}$ \\
$a_2$ & $-240.46 + 58.405y - 9.8556y^{2} + 3.1401y^{3} - 0.88932y^{4}$ \\
$a_3$ & $0.49935 + 0.60919y + 0.0024963y^{2} - 0.0099910y^{3}$ \\
$a_4$ & $2.5094(y + 3.32) + 4.1350(y + 3.32)^{2} - 0.89534(y + 3.32)^{3} - 0.0027577(y + 3.32)^{4} + 0.014511(y + 3.32)^{5}$ \\
$a_5$ & $8.2046\tpow{-7} + 1.4085\tpow{-6}\log_{10}{(0.016793(y + 4.3))} + 0.00013340/(y + 4.7136)^{2}$ \\
$a_6$ & $-267.55 - 0.21018\log_{10}{(0.35217(y + 3.9))} + 69.586y - 9.9930y^{2}$ \\
$a_7$ & $2742.8 + 222.01\log_{10}{(9.7401(y + 3.9))} - 4772.5/(y + 19.773) - 6.1001y^{2}$ \\
$a_8$ & $-0.11857 + 0.39072y - 0.037813y^{2} + 0.0022265y^{3} + 0.0046931y^{4}$ \\
\hline
$r(y)$ & $2.23\exp{(-93.4((y + 3.25)/(1 + 8.38(y + 3.25)))^{2})} - 0.376y - 0.121y^{2}$ for $T_{p}\leq 15.6$ GeV \\
           & 1.0 for $T_{p} > 15.6$ GeV \\
\hline
\multicolumn{2}{l}{Diffraction, equation \ref{eq:Fdiff}} \\
\hline 
$b_0$ & $64.682\tanh{(-0.34313(y + 2.2))} - 5.5955(y + 0.44754)^{2} + 0.0050117(y + 9.9165)^{4}$ \\
$b_1$ & $-7.6016 + 3.0427\tpow{4}\exp{(-1.0134\tpow{4}((y + 2.3066)/(1.0 + 41.612(y + 2.3066)))^{2})}$ \\
$b_2$ & $-1.4978 - 0.58163\tanh{(-0.36488(y + 1.9))} + 0.031825(y + 2.8097) + 0.022796(y - 1.8861)^{2}$ \\
$b_3$ & $-0.0061483 - 65.799\exp{(-4.8239((y + 3.8835)/(1.0 + 0.53343(y + 3.8835)))^2)}$ \\
      & Important: $b_{0},\ldots,b_{3}$ = 0 for $T_{p} < 11.05$ GeV \\
$b_4$ & $2.8009 + 0.35341(y + 2.95)^{2} - 0.0039779(y + 2.95)^{4} + 1.3012\exp{(-10.592(y + 2.2 - 0.19149(y + 2.2)^{2})^{2})}$ \\
$b_5$ & $1.8016 - 0.69847\tanh{(2.8627(y + 1.9))} - 0.015722*(y - 45.4)$ \\
$b_6$ & $1.4617 + 1.0167y - 0.078617y^{2} - 0.0038336y^{3} + 0.010141y^{4}$ \\
$b_7$ & $3.5599 + 4.0041\tanh{(-0.41889(y + 2.1))} - 0.18182*(y - 2.4209)$ \\
\hline
\multicolumn{2}{l}{$\Delta$(1232), equation \ref{eq:Fdelta}} \\
\hline
$c_0$ & $3.6052\exp{(-60.914((y + 3.1278)/(1.0 - 0.19497(y + 3.1278)))^{2})} - 0.92514 + 2.1315/y + 0.23548y^{2}$ \\
$c_1$ & $95.310 + 70.497y + 13.636y^{2}$ \\
$c_2$ & $-6.2158 - 6.2939\tanh(21.592(y + 2.1)) + 0.37440y$ \\
$c_3$ & $2.7485 + 1.1692y$ \\
$c_4$ & $-2.7568 - 1.8461y - 0.31376y^{2}$ \\
\hline
\multicolumn{2}{l}{Res1600), equation \ref{eq:Fdelta}} \\
\hline
$d_0$ & $2.5489\exp{(-58.488((y + 2.9509)/(1.0 + 1.3154(y + 2.9509)))^{2})} - 0.83039 - 0.34412y$ \\
$d_1$ & $88.173 + 65.148y + 12.585y^{2}$ \\
$d_2$ & $-7.0962 - 7.1690\tanh(30.890(y + 2.1)) + 0.38032y$ \\
$d_3$ & $-4.1440 - 3.2717y - 0.70537y^{2}$ \\
$d_4$ & $2.2624 + 1.1806y - 0.0043450y^{2} - 0.043020y^{3}$ \\
\hline\hline
\end{tabular}}
\end{table}

\begin{table}
\caption{Parameters describing muon anti-neutrino spectra for arbitrary proton energy.}
\label{table:antinumuParameters}
\scalebox{0.75}{\begin{tabular}{ll}
\hline\hline
Parameters & Formulae as functions of the proton kinetic energy 
($y=\log_{10}(T_{p})$) in TeV. \\
\hline
\multicolumn{2}{l}{Non-diff., equation \ref{eq:Fnd}} \\
\hline
$a_0$ & $-1.5243(y + 3.3) + 10.107(y + 3.3)^{2} - 4.3126(y + 3.3)^{3} + 0.80081(y + 3.3)^{4} - 0.048724(y + 3.3)^{5}$ \\
$a_1$ & $-2.6297\tpow{-5} + 9.3858\tpow{-5}\exp{(-0.32384(y + 3.4))} + 7.7821\tpow{-6}/(y + 4.0560) + 7.6149\tpow{-6}y - 8.4091\tpow{-6}y^{2}$ \\
$a_2$ & $-223.62 + 59.374y - 5.7356y^{2} + 1.9815y^{3} - 1.0478y^{4}$ \\
$a_3$ & $0.50807 + 0.60221y + 0.0034120y^{2} - 0.011139y^{3}$ \\
$a_4$ & $2.6483(y + 3.32) + 4.4585(y + 3.32)^{2} - 1.2744(y + 3.32)^{3} - 0.11659(y + 3.32)^{4} + 0.0030477(y + 3.32)^{5}$ \\
$a_5$ & $9.1101\tpow{-7} + 1.3880\tpow{-6}\log_{10}{(0.016998(y + 4.3))} 1.2744\tpow{-4}/(y + 4.7707)^{2}$ \\
$a_6$ & $-272.11 - 53.477\log_{10}{(0.35531(y + 3.9))} + 56.041/(y - 6.0876)^{2}$ \\
$a_7$ & $6431.8 + 893.92\log_{10}{(5.713\tpow{-9}(y + 3.9))} + 2103.6/(y + 5.6740) - 6.1125y^{2}$ \\
$a_8$ & $-0.11120 + 0.38144y - 0.040128y^{2} + 0.0047484y^{3} + 0.0054707y^{4}$ \\
\hline
$r(y)$ & $2.56\exp{(-107((y + 3.25)/(1 + 8.34(y + 3.25)))^{2})} - 0.385y - 0.125y^{2}$ for $T_{p}\leq 15.6$ GeV \\
           & 1.0 for $T_{p} > 15.6$ GeV \\
\hline
\multicolumn{2}{l}{Diffraction, equation \ref{eq:Fdiff}} \\
\hline 
$b_0$ & $70.430\tanh{(-0.35816(y + 2.2))} - 6.6796(y + 0.52273)^{2} + 0.0065659(y + 9.5266)^{4}$ \\
$b_1$ & $-8.1145 + 7686.0\exp{(4.4046\tpow{4}((y + 2.2190)/(1.0 + 81.105(y + 2.2190)))^{2})}$ \\
$b_2$ & $-1.3095 + 0.071270\tanh{(-0.0075463(y + 1.9))} + 0.067759(y + 5.3433) - 0.0044205(y - 1.8683)^{2}$ \\
$b_3$ & $0.082149 - 2190.1\exp{(-533.75((y + 2.8363)/(1.0 + 7.0976(y + 2.8363)))^2)}$ \\
      & Important: $b_{0},\ldots,b_{3}$ = 0 for $T_{p} < 11.05$ GeV \\
$b_4$ & $2.7540 + 0.33859(y + 2.95)^{2} - 0.0034274(y + 2.95)^{4} + 1.1679\exp{(-10.408(y + 2.2 - 0.18922(y + 2.2)^{2})^{2})}$ \\
$b_5$ & $2.1817 - 0.59584\tanh{(2.7054(y + 1.9))} - 0.010909*(y - 14.9)$ \\
$b_6$ & $1.4591 + 1.0275y - 0.074949y^{2} - 0.0060396y^{3} + 0.0097568y^{4}$ \\
$b_7$ & $3.7609 + 4.2843\tanh{(-0.37148(y + 2.1))} - 0.16479*(y - 2.7653)$ \\
\hline
\multicolumn{2}{l}{$\Delta$(1232), equation \ref{eq:Fdelta}} \\
\hline
$c_0$ & $2.8262\exp{(-62.894((y + 3.1250)/(1.0 - 0.47567(y + 3.1250)))^{2})} + 5.6845 + 13.409/y - 0.097296y^{2}$ \\
$c_1$ & $16.721 + 11.750y + 2.4637y^{2}$ \\
$c_2$ & $-6.0557 - 6.3378\tanh(21.984(y + 2.1)) + 0.43173y$ \\
$c_3$ & $0.37009 + 0.27706y$ \\
$c_4$ & $0.047507 + 0.061570y + 0.0070117y^{2}$ \\
\hline
\multicolumn{2}{l}{Res1600), equation \ref{eq:Fdelta}} \\
\hline
$d_0$ & $2.2400\exp{(-57.159((y + 2.9492)/(1.0 + 1.2994(y + 2.9492)))^{2})} - 0.66521 - 0.27554y$ \\
$d_1$ & $-7.0650 - 4.2773y - 0.17648y^{2}$ \\
$d_2$ & $-7.0410 - 7.1977\tanh(31.095(y + 2.1)) + 0.40238y$ \\
$d_3$ & $-1.2354 - 0.87581y - 0.20829^{2}$ \\
$d_4$ & $-0.11395 + 0.34418y + 0.27103y^{2} + 0.050248y^{3}$ \\
\hline\hline
\end{tabular}}
\end{table}

\end{document}